\documentclass[prl,twocolumn,superscriptaddress,amsmath]{revtex4-1}

\usepackage{graphicx}
\usepackage{amsmath,amssymb}
\usepackage{times}
\usepackage{comment}
\usepackage[colorlinks=true,linkcolor=blue,citecolor=blue]{hyperref}
\newcommand{\eqn}[1]{
\begin{eqnarray}
	#1
\end{eqnarray}
}

\begin{document}

\title{Parity-time-symmetric quantum critical phenomena}

\author{Yuto Ashida}
%\email{ashida@cat.phys.s.u-tokyo.ac.jp}
\affiliation{Department of Physics, University of Tokyo, 7-3-1 Hongo, Bunkyo-ku, Tokyo
113-0033, Japan}
\author{Shunsuke Furukawa}
\affiliation{Department of Physics, University of Tokyo, 7-3-1 Hongo, Bunkyo-ku, Tokyo
113-0033, Japan}
\author{Masahito Ueda}
%\email{ueda@phys.s.u-tokyo.ac.jp}
\affiliation{Department of Physics, University of Tokyo, 7-3-1 Hongo, Bunkyo-ku, Tokyo
113-0033, Japan}
\affiliation{RIKEN Center for Emergent Matter Science (CEMS), Wako, Saitama 351-0198, Japan
}

\date{\today}

%\begin{abstract}
%\end{abstract}
\nopagebreak

\begin{abstract}
Synthetic nonconservative systems with parity-time (PT) symmetric gain-loss structures can exhibit unusual spontaneous symmetry breaking that accompanies spectral singularity. Recent studies on PT symmetry in optics and weakly interacting open quantum systems have revealed  intriguing physical properties, yet many-body correlations still play no role. Here by extending the idea of PT symmetry to  strongly correlated many-body systems, we report that a combination of spectral singularity and quantum criticality  yields an exotic universality class which has no counterpart in known critical phenomena. Moreover, we find unconventional low-dimensional quantum criticality, where superfluid correlation is anomalously enhanced owing to non-monotonic renormalization group flows in a PT-symmetry-broken quantum critical phase, in stark contrast to the Berezinskii-Kosterlitz-Thouless paradigm. Our findings can be experimentally tested in ultracold atoms and predict critical phenomena beyond the Hermitian paradigm of quantum many-body physics.
\end{abstract}

\maketitle

Studies of phase transitions and critical behaviour in non-Hermitian systems date back to the discovery of the Lee-Yang edge singularity \cite{MEF78}, where an imaginary magnetic field in the high-temperature Ising model was demonstrated to trigger an exotic phase transition. More recently, the real-to-complex spectral phase transition has been found in a broad class of non-Hermitian Hamiltonians that satisfy parity-time (PT) symmetry \cite{CMB98}. While such systems were once of purely academic interest, related questions are now  within experimental reach \cite{CER10,AR12,LF13,BP14,BZ15}.

 A Hamiltonian $\hat{H}$ is said to be PT-symmetric if it commutes with the combined operator $\hat{P}\hat{T}$, but not necessarily with $\hat{P}$ and $\hat{T}$ separately. Here $\hat{P}$ and $\hat{T}$ are the parity and time-reversal operators, respectively. The PT symmetry is said to be unbroken if every  eigenstate of $\hat{H}$ is PT-symmetric; then, the entire spectrum is real even though $\hat{H}$ is not Hermitian. The PT symmetry is said to be  spontaneously broken if some   eigenstates of $\hat{H}$ are not the eigenstates of the PT operator even though $[\hat{H},\hat{P}\hat{T}]=0$; then, some pairs of eigenvalues of $\hat{H}$ become complex conjugate to each other. The PT  symmetry breaking is typically accompanied by the coalescence of  eigenstates and that of the corresponding eigenvalues at  an exceptional point \cite{TK80} in the discrete spectrum or the spectral singularity \cite{AM09} in the continuum spectrum. 
 While these features also hold for a certain class of antilinear symmetries \cite{CMB02}, PT symmetry allows experimental implementations by spatial engineering of gain-loss structures, leading to a rich interplay between theory and experiment in optics \cite{KGM08,AR12,LF13,BP14,BZ15}, superconductors \cite{NMC12}, atomic physics \cite{PP16}, and optomechanics \cite{HJ14}. In particular, the real-to-complex spectral transition (PT transition) has been observed in experiments of classical systems \cite{CMB13}. In all these developments, however, many-body correlations still play no role.
 
Quantum critical phenomena, in contrast, arise from collective behaviour of strongly correlated systems and exhibit universal long-distance properties. In view of recent developments in designing open many-body systems in ultracold atoms \cite{BWS09,BG13,BF13,PYS15,YA15} and exciton-polariton condensates \cite{GT15}, it seems ripe to explore the role of PT symmetry in quantum critical phenomena and ask whether or not the concept of the universality need be extended in synthetic nonconservative systems.

Here we report that a combination of spectral singularity and quantum criticality yields an exotic critical point in the extended parameter space and that, in the PT-broken phase, a local gain-loss structure results in an anomalous enhancement of superfluid correlation owing to semicircular renormalization group  (RG) flows. This contrasts sharply with the suppression of superfluid correlation due to hyperbolic RG flows in the Berezinskii-Kosterlitz-Thouless (BKT) paradigm. Our findings demonstrate that the interplay between many-body correlations and PT symmetry leads to the emergence of quantum critical phenomena beyond the Hermitian paradigm of quantum many-body physics.
\\
\\
\noindent
{\bf Results}
\\
\noindent
{\bf Parity-time-symmetric sine-Gordon model.}
We consider a class of one-dimensional (1D) quantum systems described by the field-theory Hamiltonian 

\eqn{\label{mH}
\hat{H}=\int dx\left\{\frac{\hbar v}{2\pi}\left[K(\partial_{x}\hat{\theta})^2+\frac{1}{K}(\partial_{x}\hat{\phi})^2\right]+V(\hat{\phi})\right\},
}
where $\hat{\phi}$ is a scalar field, $\partial_x\hat{\theta}$ is its conjugate momentum 
satisfying $[\hat{\phi}(x),\partial_x\hat{\theta}(x')]=-i\pi\delta(x-x')$, 
and $V(\hat{\phi})$ is a potential for the field $\hat{\phi}$. 
Without the potential term, equation (1) is known as the Tomonaga-Luttinger liquid (TLL) Hamiltonian, which gives a universal framework for describing 1D-interacting bosons and fermions \cite{TG03}.  
Here, $v$ is the sound velocity, the TLL parameter $K$ characterizes the interaction strength, 
$\partial_{x}\hat{\phi}$ and $\hat{\theta}$ are related to the density and the Josephson phase, respectively. 
The introduction of the cosine potential $V(\hat{\phi}) \propto \cos(2\hat{\phi})$ results in the sine-Gordon model, 
which describes the BKT transition to a gapped phase. 
For bosons on a lattice, this corresponds to a superfluid-to-Mott-insulator (MI) transition \cite{FM89}. 
Here we consider a generalization to the PT-symmetric case 
by adding an imaginary contribution to the potential term as follows: 

\eqn{\label{ptp}
V(\hat{\phi})=\frac{\alpha_{\rm r}}{\pi}\cos(2\hat{\phi})-\frac{i \alpha_{\rm i}}{\pi}\sin(2\hat{\phi}),
}
where $\alpha_{\rm r}$ and $\alpha_{\rm i}$ characterize the strengths of the real and imaginary parts of the potential. When the real part becomes relevant, it suppresses the fluctuations of $\hat{\phi}$, stabilizing a non-critical, gapped phase. In contrast, if the imaginary part is relevant, it facilitates the fluctuations of $\hat{\phi}$ and enhances correlation in the conjugate field $\hat{\theta}$, as we will see later.
The field theory (\ref{mH}) with the potential (\ref{ptp}) satisfies PT symmetry since the field $\hat{\phi}$  has odd parity. The PT-symmetric Hamiltonian $\hat{H}$ can be implemented by a continuously monitored 1D-interacting ultracold atoms (see Supplementary Note 1 and Supplementary Figure 1).

\begin{figure}
\includegraphics[width=86mm]{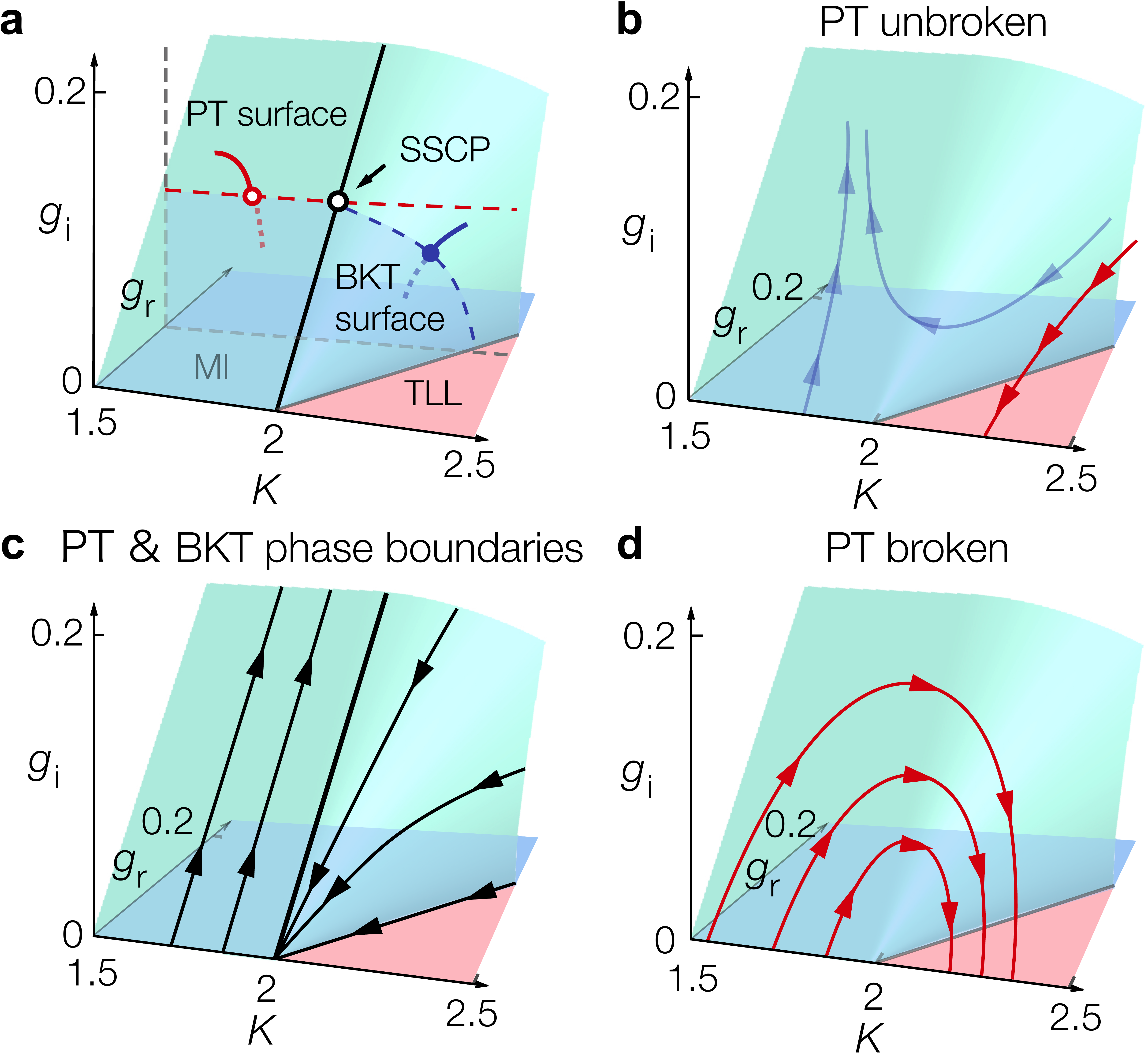} 
\caption{\label{fig1}
{\bf Quantum critical phenomena in PT-symmetric many-body systems.}  ({\bf a}) Three-dimensional phase diagram of a PT-symmetric many-body system in the parameter space ($K,g_{\rm r},g_{\rm i}$). Here $K$ and $g_{\rm r}$ ($g_{\rm i}$) characterize the strength of the inter-particle interaction and the depth of the real (imaginary) part of a complex potential, respectively. The Mott insulator (MI) and Tomonaga-Luttinger liquid (TLL) phases are separated by the surface of the Berezinskii-Kosterlitz-Thouless (BKT) transition for $K>2$ and that of the PT transition for $K<2$. An example of the BKT (PT) transition is illustrated by the blue (red) curve with the transition point indicated by the filled (open) circle. The MI (TLL) phase corresponds to the three-dimensional region containing the blue (red) shaded plane at $g_{\rm i}=0$. On the critical line with $K=2$ lies a spectral singular critical point (SSCP, black open circle). Dashed lines indicate the phase boundaries on the plane with fixed $g_{\rm r}$ for comparison with numerical results in Fig.~\ref{fig2}a. ({\bf b}) Hyperbolic renormalization group (RG) flows in a PT-unbroken region $(g_{\rm i}<g_{\rm r})$, which reproduce the conventional flow diagram in the sine-Gordon model. ({\bf c}) RG flows on the two phase boundaries separated by an unconventional fixed line (thick black line). ({\bf d}) Unconventional semicircular RG flows in a PT-broken region $(g_{\rm i}>g_{\rm r})$. Along each flow, the TLL parameter $K$ monotonically increases, indicating the anomalous enhancement of the superfluid correlation.
}
\end{figure}

We note that if $\alpha_{\rm r}>\alpha_{\rm i}$,  $\hat{H}$ has a real spectrum and thus PT symmetry is unbroken. This can be proved by the theorem \cite{AR02} which states that the spectrum is real if and only if there exists an operator $\hat{O}$ satisfying $\hat{O}^{-1}\hat{H}\hat{O}=\hat{H}_{0}$, where $\hat{H}_{0}$ is a Hermitian operator. Indeed, we can explicitly construct such an operator for $\alpha_{\rm r}>\alpha_{\rm i}$ by the choice of $\hat{O}= e^{-\eta\hat{\theta}_{0}/2}$, where $\hat{\theta}_{0}$ is a constant part of $\hat{\theta}$ and $\eta\equiv{\rm arctanh} (\alpha_{\rm i}/\alpha_{\rm r})$. Then, the potential term in the effective field theory is transformed to $(\sqrt{\alpha_{\rm r}^2-\alpha_{\rm i}^2}/\pi)\int dx \cos(2\hat{\phi})$ and $\hat{H}$ reduces to the sine-Gordon Hamiltonian \cite{CMB05}. Divergence of $\eta$ at $\alpha_{\rm r}=\alpha_{\rm i}$ signals spontaneous  breaking of PT symmetry.
\\
\\
\noindent
{\bf Renormalization group analysis.}
To unravel the universal critical behaviour of the PT-symmetric Hamiltonian $\hat{H}$, we perform an RG analysis \cite{DJA80} to obtain the following set of flow equations which are valid up to the third order in $g_{\rm r,i}$:
\eqn{
\frac{{\rm d}K}{{\rm d}l}&=&-(g_{\rm r}^2-g_{\rm i}^2) K^{2},\;\;\;\;\frac{{\rm d}g_{\rm r}}{{\rm d}l}=(2-K)g_{\rm r}+5g_{\rm r}^3-5g_{\rm i}^2 g_{\rm r},\nonumber\\
\frac{{\rm d}g_{\rm i}}{{\rm d}l}&=&(2-K)g_{\rm i}-5g_{\rm i}^3+5g_{\rm r}^2 g_{\rm i}.
}
Here $l$ is the logarithmic RG scale and $g_{\rm r,i}\equiv\alpha_{\rm r,i}a^{2}/(\hbar v)$ are the dimensionless coupling constants with $a$ being a short-distance cutoff. The velocity $v$ stays constant to all orders in $g_{\rm r,i}$ because of the Lorentz invariance of the theory. 
In contrast to the two-dimensional phase diagram of the conventional sine-Gordon model, the PT-symmetric system has the three-dimensional phase diagram (Fig.~\ref{fig1}a). When PT symmetry is unbroken, i.e., $g_{\rm i}<g_{\rm r}$, the spectrum is equivalent to that of the closed system as discussed above and the conventional RG flow diagram with hyperbolic flows is reproduced  (Fig.~\ref{fig1}b). Here the BKT boundary between the superfluid TLL phase and the MI phase extends over the curved surface. We note that the operator $\hat{O}$ does not affect the critical properties of the ground state since it only modifies the zero modes associated with the field $\hat{\phi}$. Since the non-Hermitian term can arise from the measurement backaction, the quantum phase transition induced by increasing $g_{\rm i}$ may be regarded as measurement-induced.

In the strongly correlated regime $K<2$, a new type of quantum phase transition appears on the PT threshold plane $g_{\rm i}=g_{\rm r}$. This phase transition is accompanied by spontaneous breaking of the PT symmetry in eigenstates, contrary to the ordinary BKT transition exhibiting no symmetry breaking. The BKT and PT phase boundaries merge on the  line defined by $K=2$ and $g_{\rm i}=g_{\rm r}$ (Fig.~\ref{fig1}c). In general, at the PT symmetry breaking point, the spectral singularity \cite{AM09} arises where two or more eigenvalues as well as their eigenstates coalesce in the continuum spectrum. In optics, the spectral singularity leads to unidirectional wave phenomena \cite{LF13}. In contrast, in many-body systems, the coexistence of the spectral singularity and the quantum criticality at $g_{\rm i}=g_{\rm r}$ and $K=2$ results in what we term a spectral singular critical point (SSCP), which represents a unique universality class in nonconservative systems.

\begin{figure}
\includegraphics[width=86mm]{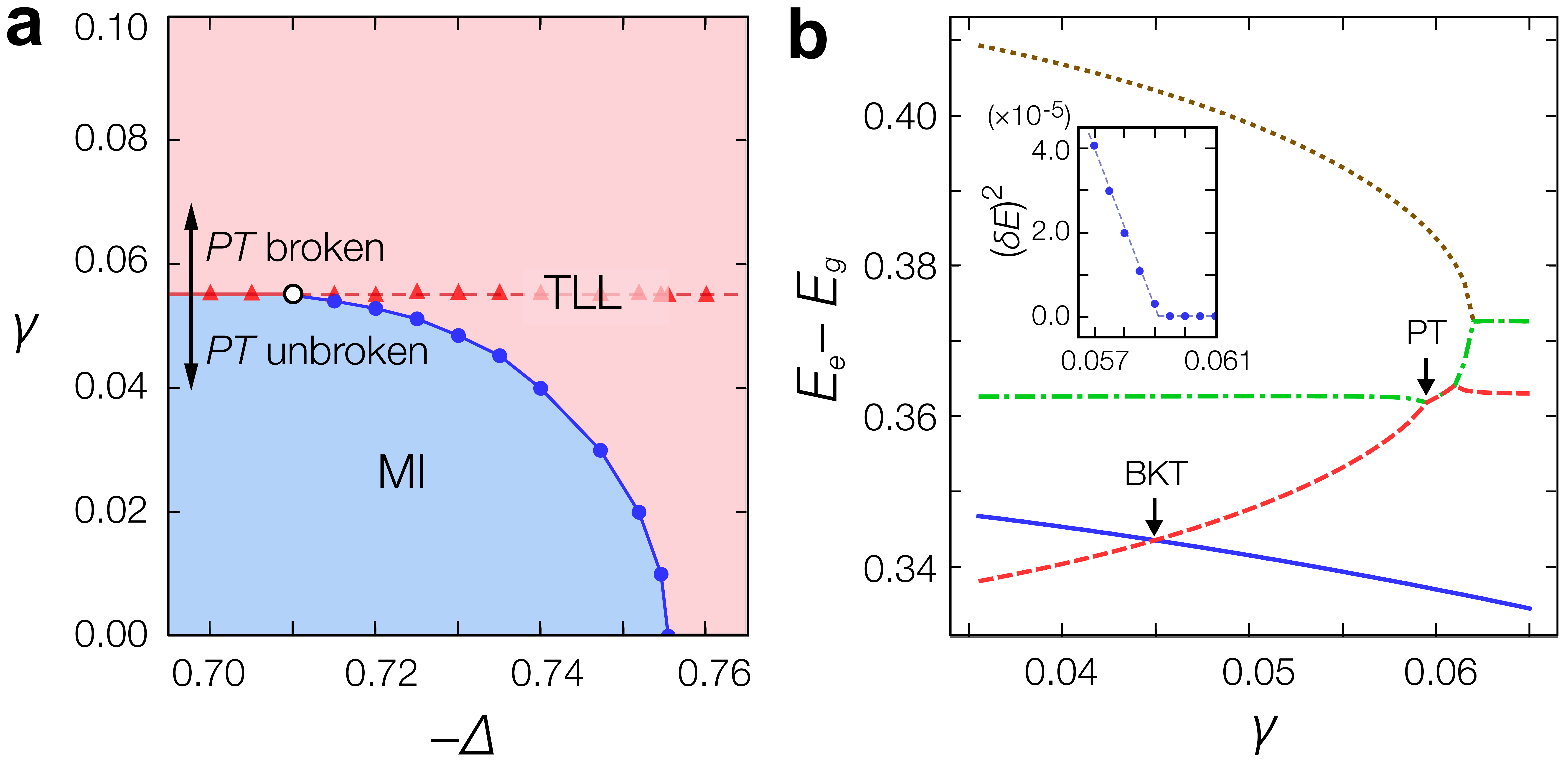}
\caption{\label{fig2}
{\bf Phase diagram and finite-size spectrum.}  ({\bf a}) Ground-state phase diagram of the PT-symmetric many-body lattice Hamiltonian. The Mott insulator (MI) and Tomonaga-Luttinger liquid (TLL) phases are separated by the Berezinskii-Kosterlitz-Thouless (BKT) transition (blue curve with filled circles) and PT-symmetry breaking (red line with filled triangles). The point where the two boundaries merge defines the spectral singular critical point (SSCP, open circle). ({\bf b}) Typical low-energy excitation spectrum in the lattice model. The three lowest levels in the $S^{z}=0$ sector (red, green, and yellow curves from the lowest), and the lowest excitation energy to the $S^{z}=\pm4$ sector (blue curve) are plotted. Here $S^{z}=\sum_{m=1}^{N}\hat{S}^{z}_{m}$ is a total magnetization. The energy difference $\delta E$ between the two coalescing levels (e.g., red and green) obeys the square-root scaling (inset) and closes at the PT-symmetry breaking point.  The BKT transition point corresponds to a crossing of appropriate levels (red and blue). We set the parameter $h_{s}=0.1$ for both ({\bf a}) and ({\bf b}). In ({\bf a}), the plotted data are obtained through extrapolation to the thermodynamic limit, while the data in ({\bf b})  are obtained for $N=16$ and $-\Delta=0.735$. The plotted variables are dimensionless since we set $J=1$.
}
\end{figure} 

\begin{figure}
\includegraphics[width=86mm]{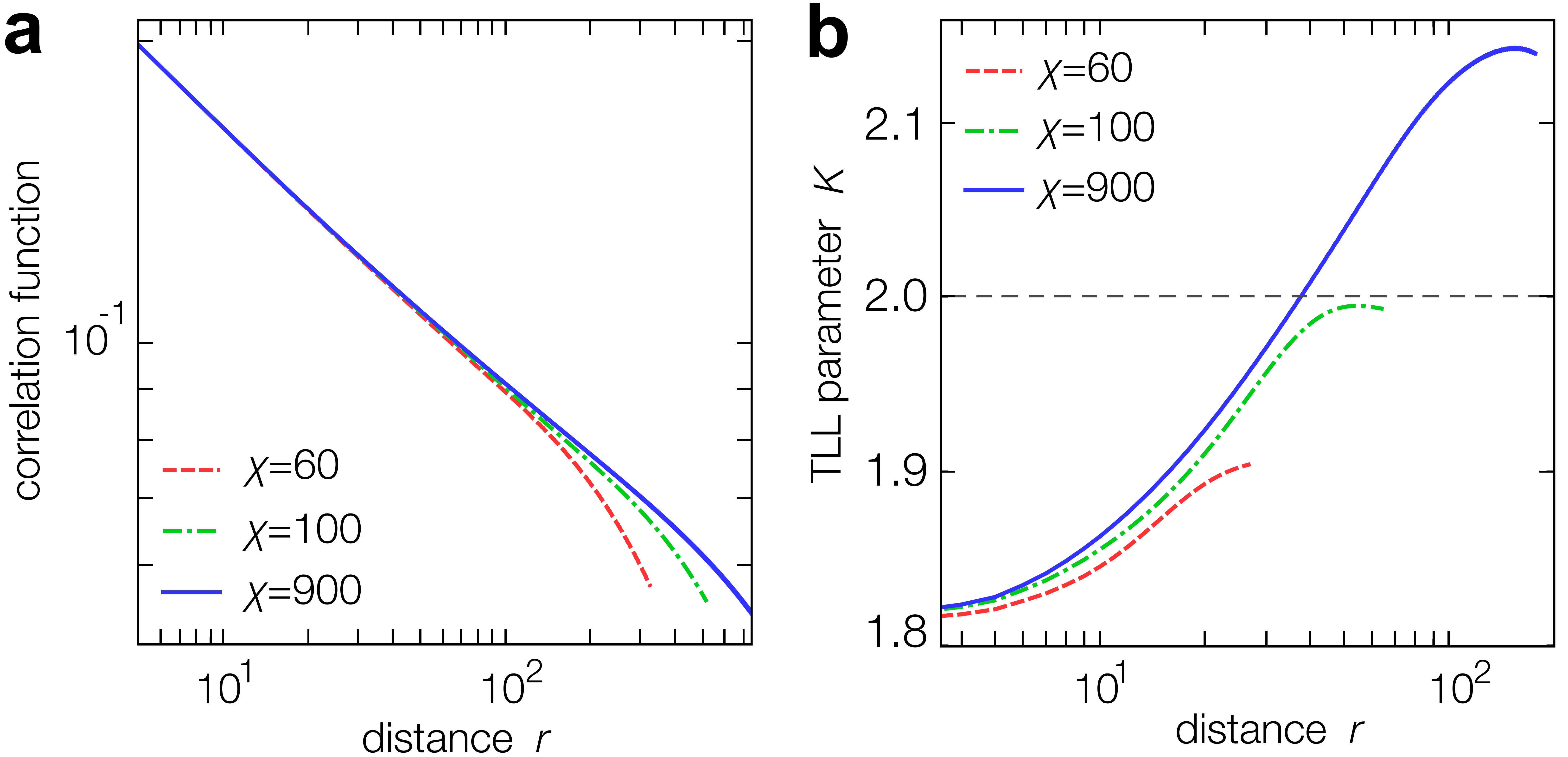}
\caption{\label{fig3}
{\bf Anomalous enhancement of superfluid correlation in the PT-broken quantum critical phase.} ({\bf a}) Critical decay of the correlation function ${\rm Re}[\langle\hat{S}^{+}_{r}\hat{S}^{-}_{0}\rangle]$. ({\bf b}) Tomonaga-Luttinger liquid (TLL) parameter $K$ as a function of the distance $r$, giving the critical exponent of the correlation function, $\langle\hat{S}^{+}_{r}\hat{S}^{-}_{0}\rangle\propto(1/r)^{1/(2K)}$. The exponent is extracted by the linear fitting of the correlation function in the log-log plot around the distance $r$. The parameters are set to be $-\Delta=0.61$, $h_{s}=0.1$, and $\gamma=0.08$, and $\chi$ denotes the dimension of the matrix product state that controls the accuracy of the iTEBD simulation. 
}
\end{figure}

\begin{figure}
\includegraphics[width=86mm]{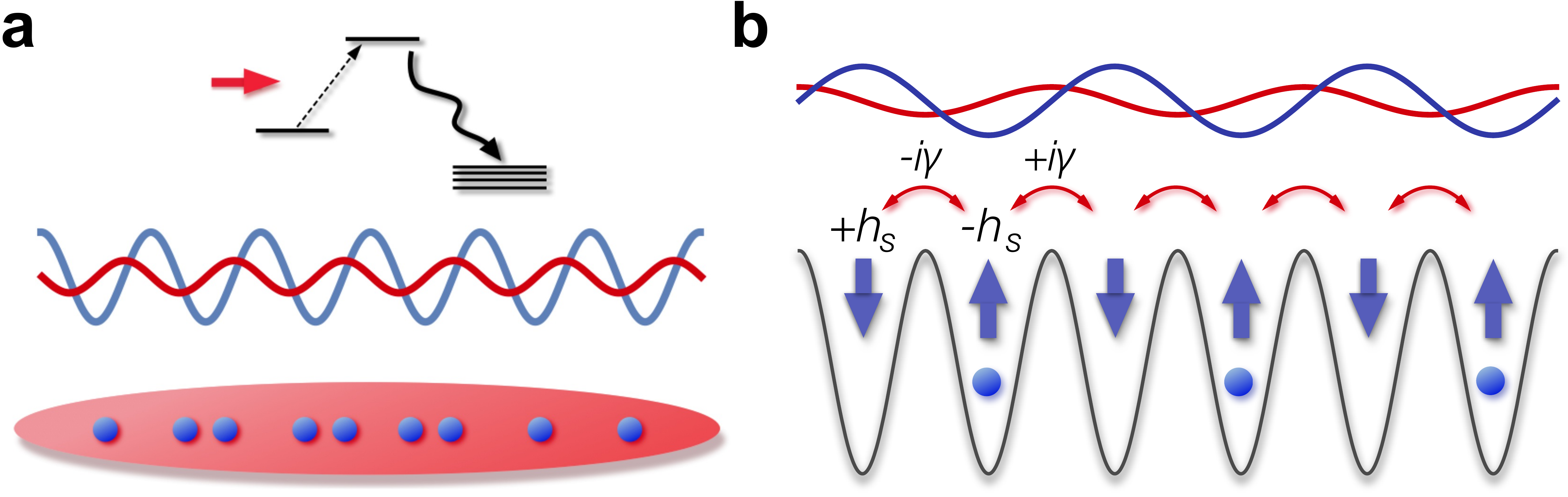}
 \caption{\label{fig4}
 {\bf Experimental setup of a PT-symmetric many-body system in ultracold atoms.} ({\bf a}) One-dimensional ultracold atoms in a PT-symmetric optical lattice. Real (blue  curve) and imaginary (red curve) parts of a complex potential are created by a pair of far-detuned and weak near-resonant standing  waves. An imaginary potential results from  a near-resonant light (red arrow) on atoms whose excited state has fast decay modes. The two periodic potentials are displaced from each other by one half of the lattice spacing so that the system possesses PT symmetry. 
({\bf b}) Mapping to a PT-symmetric  lattice model that reproduces the same critical behaviour as the continuum model. Atoms are strongly localized by a deep optical lattice that does not affect the universal critical behaviour. The real and the imaginary parts of the complex potential introduce the on-site potentials $\pm h_{\rm s}$ and imaginary hopping terms $\pm i\gamma$. A lattice site occupied (not occupied) by a hard-core boson is represented by the up (down) spin.
}
\end{figure}

When the PT symmetry is broken, i.e., $g_{\rm i}>g_{\rm r}$,  unconventional RG flows emerge: starting from the $K<2$ side, $g_{\rm r,i}$ and $K$ initially increase, and after entering the $K>2$ side, the flow winds and converges to the fixed line with $g_{\rm r,i}=0$ (Fig.~\ref{fig1}d). Physically, this significant increase in the TLL parameter $K$ indicates that the superfluid correlation decays more slowly and is thus enhanced by the non-Hermiticity of an imaginary potential. The enhancement is viewed as anomalous because, in the conventional BKT paradigm, a real potential suppresses the fluctuation of $\hat{\phi}$ and stabilizes the gapped MI phase for $K<2$. Moreover, owing to the semicircular RG flows, the imaginary potential allows for a substantial increase of the TLL parameter $K$ even if its strength $g_{\rm i}$ is initially very small. The PT-broken phase exhibits other observable consequences such as anomalous lasing and absorption as observed in optics \cite{PB14} (see Supplementary Note 2 for the experimental implementation in ultracold atoms).
\\
\\
{\bf Ground-state phase diagram of the lattice model.}
To numerically demonstrate these findings, we introduce a lattice Hamiltonian
\eqn{
\hat{H}_{\rm L}&=&\sum_{m=1}^{N}\Bigl[-\left(J+(-1)^{m}i\gamma\right)\left(\hat{S}^{x}_{m}\hat{S}^{x}_{m+1}+\hat{S}^{y}_{m}\hat{S}^{y}_{m+1}\right)\nonumber\\
&&\;\;\;\;\;\;\;\;\;+\Delta\hat{S}^{z}_{m}\hat{S}^{z}_{m+1}+(-1)^{m}h_{\rm s}\hat{S}^{z}_{m}\Bigr],\label{HL}
}
whose low-energy behaviour is described by the PT-symmetric effective field theory $\hat{H}$. 
Here $\hat{S}^{x,y,z}_{m}$ are the spin-1/2 operators at site $m$ and the parameters $(-\Delta,h_{\rm s},\gamma)$ are related to $(K,g_{\rm r},g_{\rm i})$ in the field theory, where we set $J=1$. The non-Hermitian term represents a periodic gain-loss structure and effectively strengthens the amplitude of the hopping term, leading to enhanced superfluid correlation. The determined phase diagram and a typical exact finite-size spectrum are shown in Fig.~\ref{fig2}. The BKT transition is identified as a crossing point of appropriate energy levels \cite{KN95} and the PT threshold is determined as a coalescence point in low-energy levels, as detailed in Methods and Supplementary Methods. The coalescence point is found to be an exceptional point from the characteristic square-root scaling \cite{TK80} of the energy gap (see the inset figure in Fig.~\ref{fig2}b). We note that, above the PT threshold, some highly excited states turn out to have positive imaginary parts of eigenvalues and cause the instability in the long-time limit. The presence of such high-energy unstable modes is reminiscent of parametric instabilities in exciton-polariton systems \cite{DS08}, and can ultimately destroy the one-dimensional coherence \cite{IC05}. In our setup, where the imaginary term is adiabatically ramped up, the amplitudes of these unstable modes can greatly be suppressed and the system can remain, with almost unit fidelity, in the ground state in which the critical behaviour is sustained (see Supplementary Note 3 and Supplementary Figures 2 and 3 for details).
\\
\\
{\bf Numerical demonstration of enhanced superfluid correlation.}
To demonstrate the anomalous enhancement of superfluid correlation in the PT-broken regime, we have performed numerical simulations using the infinite time-evolving block decimation (iTEBD) algorithm \cite{GV07}.  The correlation function exhibits the critical decay with a varying critical exponent and the corresponding TLL parameter significantly increases, surpassing $K=2$ as shown in Fig.~\ref{fig3}. Physically, this enhancement of superfluid correlation at long distances can be interpreted as follows. A local gain-loss structure introduced by the imaginary term causes locally equilibrated flows \cite{CMB13} in the ground state. This results in the enhancement of fluctuations in the density, or equivalently, the suppression of fluctuations in the conjugate phase. It is this effect that increases the superfluid correlation. The numerical results are consistent with the analytical arguments given above, and demonstrate that the RG analysis is instrumental in studying critical properties of a non-Hermitian many-body system. 
\\
\\
{\bf Experimental realization in a one-dimensional Bose gas.}
The PT-symmetric many-body Hamiltonian $\hat{H}$ discussed above can be implemented in  a 1D-interacting ultracold bosonic atoms subject to a shallow PT-symmetric optical lattice $V(x)=V_{\rm r}\cos(2\pi x/d)-iV_{\rm i}\sin(2\pi x/d)$, where $V_{\rm r}$ and $V_{\rm i}$ are the depths of the real and imaginary parts of a complex potential and $d$ is the lattice constant. An imaginary optical potential can be realized by using a weak near-resonant standing-wave light (see  Methods). 
Since $V(x)$ remains invariant under simultaneous parity operation ($x\to-x$) and time reversal (i.e., complex conjugation), the system satisfies the condition of PT symmetry (Fig.~\ref{fig4}a).
In open quantum systems, by postselecting null measurement outcomes, the time evolution is governed by an effective non-Hermitian Hamiltonian \cite{HC93,AD15,YA16}. The achieved experimental fidelity has already been high enough to allow experimenters to implement various types of postselections \cite{EM11,FT15,RI15}.  The low-energy behaviour of this system is then described by the PT-symmetric effective field theory $\hat{H}$. We note that the lattice Hamiltonian (\ref{HL}) can also be realized in ultracold atoms by superimposing a deep lattice that does not influence the universal critical behaviour (see Fig.~\ref{fig4}b).

We stress that the dynamics considered here is different from the one described by a master equation, where dissipative processes,  in general, tend to destroy correlations underlying quantum critical phenomena. In contrast, the postselections allow us to study the system free from the dissipative jump processes, while non-trivial effects due to measurement backaction still occur via the non-Hermitian contributions in the effective Hamiltonian.
\\
\\
{\bf Discussion}
The reported fixed points in the extended parameter space suggest that an interplay between spectral singularity and quantum criticality results in an exotic universality class beyond the conventional paradigm.  It remains an open question how the universality accompanying spectral singularity found in this work is related to non-unitary conformal field theories (CFT) studied in various fields ranging from statistical mechanics \cite{JC85} to high-energy physics \cite{NS90}. It is particularly notable that a certain critical point of the integrable spin chain with PT-symmetric boundary fields corresponds to an exceptional point and is believed to be described by non-unitary CFT \cite{VP90}. This suggests an intimate connection between the SSCP and the non-unitary CFT. Given recent success in measuring entanglement entropy in ultracold atoms \cite{RI15}, it is of interest to study how quantum entanglement behaves in the presence of spectral singularity. 
In the PT-broken phase, we have shown that the ground state exhibits the enhanced superfluid correlation indicating the tighter binding of the topological excitations, in stark contrast to their proliferation as found in the BKT paradigm.  
In Hermitian systems, a relevant perturbation around RG fixed points has a tendency to suppress fluctuations of the concerned field and  stabilize a non-critical, gapped phase.
Our finding indicates that a relevant imaginary perturbation can realize the opposite situation of enhancing fluctuations of the concerned field and facilitating correlation in the conjugate field. An exploration of such unconventional quantum criticality in other synthetic, nonconservative many-body systems presents an interesting challenge. Further studies in these directions, together with their possible experimental realizations, could widen applications to future quantum metamaterials.
\\
\\
\noindent
{\bf Methods}
\\
\noindent
{\bf Details of numerical calculations.} The phase diagram in Fig.~\ref{fig2}a is determined from the exact diagonalization analysis of the lattice Hamiltonian (\ref{HL}). To identify the BKT transition point, we calculate the exact finite-size spectrum and find a crossing of low-energy levels having appropriate quantum numbers \cite{KN95}. The PT transition point is identified as the first  coalescence point in the low-energy spectrum with increasing $\gamma$. The calculations are done for different system sizes and the final results are obtained through extrapolation of the data to the thermodynamic limit. Further details are given in Supplementary Methods and Supplementary Figure 4.
The correlation function and the associated variation of the TLL parameter $K$ shown in Fig.~\ref{fig3} are calculated by applying the iTEBD algorithm \cite{GV07}. We emphasize that this method can be applied to study the ground-state properties of the non-Hermitian system. The method can accurately calculate the imaginary-time evolution $\exp(-\hat{H}\tau)|\Psi_{0}\rangle/\|\exp(-\hat{H}\tau)|\Psi_{0}\rangle\|$ for an infinite system size, where $\tau$ is an imaginary time, $|\Psi_{0}\rangle$ is an initial state and $\|\cdot\|$ denotes the norm of the state. In the limit of large $\tau$, we obtain the quantum state, the real part of which eigenvalue is the lowest in the entire spectrum, i.e., an  effective ground state of a non-Hermitian system. We note that the imaginary part of the eigenvalue does not affect the calculation since it only changes the overall phase of the wavefunction in the imaginary-time evolution. We then determine the TLL parameter $K$ from the calculated correlation function by using the relation $\langle\hat{S}^{+}_{r}\hat{S}^{-}_{0}\rangle\propto (1/r)^{1/(2K)}$.
\\
\\
{\bf Derivation of the low-energy field theory of ultracold atoms.} Here we explain the derivation of the low-energy effective field theory (\ref{mH}) of ultracold atoms. We start from the Hamiltonian in which the periodic  potential $V_{\rm r}\cos(2\pi x/d)$ is added to the Lieb-Liniger model \cite{EHL63}. Then, we introduce an imaginary optical lattice potential by using a weak near-resonant standing-wave light. This scheme is possible if the excited state $|e\rangle$ of an atom has decay modes other than the initial ground state $|g\rangle$ and its decay rate is faster than the spontaneous decay rate from $|e\rangle$ to $|g\rangle$ and the Rabi frequency \cite{MKO99,AT03,RS05} (Fig.~\ref{fig4}a). Such a condition can be satisfied by,  e.g., using appropriate atomic levels \cite{KSJ98} or light-induced transitions \cite{BWS09}. The difference between the wavelengths of the real and imaginary periodic potentials caused by different detunings of the lasers can be negligible. Using the second-order perturbation theory  \cite{TK80} for the Rabi coupling and adiabatically eliminating the excited state, we obtain an effective time-evolution equation for the ground-state atoms. We then assume that null measurement outcomes are postselected so that the dynamics is described by the non-Hermitian Hamiltonian \cite{HC93,AD15,YA16}. In this situation, the overall imaginary constant in the eigenvalue spectrum does not affect the dynamics since it can be eliminated when we normalize the quantum state, leading to the imaginary potential $iV_{\rm i}\sin(2\pi x/d)$.  Finally, we follow the standard procedure \cite{TG03} of taking the low-energy limit of the model and arrive at the Hamiltonian (\ref{mH}). The details of the calculations and experimental accessibility in ultracold atoms are described in Supplementary Note 1 and 2.

\vspace{1cm}
\noindent
{\bf Acknowledgements}
We acknowledge support from KAKENHI Grant Nos. JP25800225 and JP26287088 from the Japan Society for the Promotion of Science (JSPS), and a Grant-in-Aid for Scientific Research on Innovative Areas ``Topological Materials Science" (KAKENHI Grant No. JP15H05855), and the Photon Frontier Network Program from MEXT of Japan, ImPACT Program of Council for Science, Technology and Innovation (Cabinet Office, Government of Japan). We are grateful to Yusuke Horinouchi, Ryusuke Hamazaki, Zongping Gong, Shintaro Takayoshi, Yuya Nakagawa, Takeshi Fukuhara, Takashi Mori, and Hosho Katsura for valuable discussions. Y. A. acknowledges support from JSPS (Grant No. JP16J03613). 
\\
\\
{\bf Author contributions}
Y.A., S.F. and M.U. planned the project. Y.A. performed the analytical calculations. Y.A. and S.F. performed the numerical calculations. Y.A., S.F. and M.U. analyzed and interpreted the results and wrote the manuscript. 
\\
\\
{\bf Competing financial interests}
The authors declare that they have no competing financial interests.

\widetext
\pagebreak
\begin{center}
\textbf{\large Supplementary Materials}
\end{center}

\renewcommand{\theequation}{S\arabic{equation}}
\renewcommand{\thefigure}{S\arabic{figure}}
\renewcommand{\bibnumfmt}[1]{[S#1]}
\setcounter{equation}{0}
\setcounter{figure}{0}

\noindent
{\bf Supplementary Note 1 - Parity-time symmetric effective Hamiltonian in ultracold atoms}
\\
Here we describe in detail how a parity-time (PT) symmetric field theory described by equation (1) in the main text can be implemented by using ultracold atoms.

\begin{figure}[b]
\includegraphics[width=60mm]{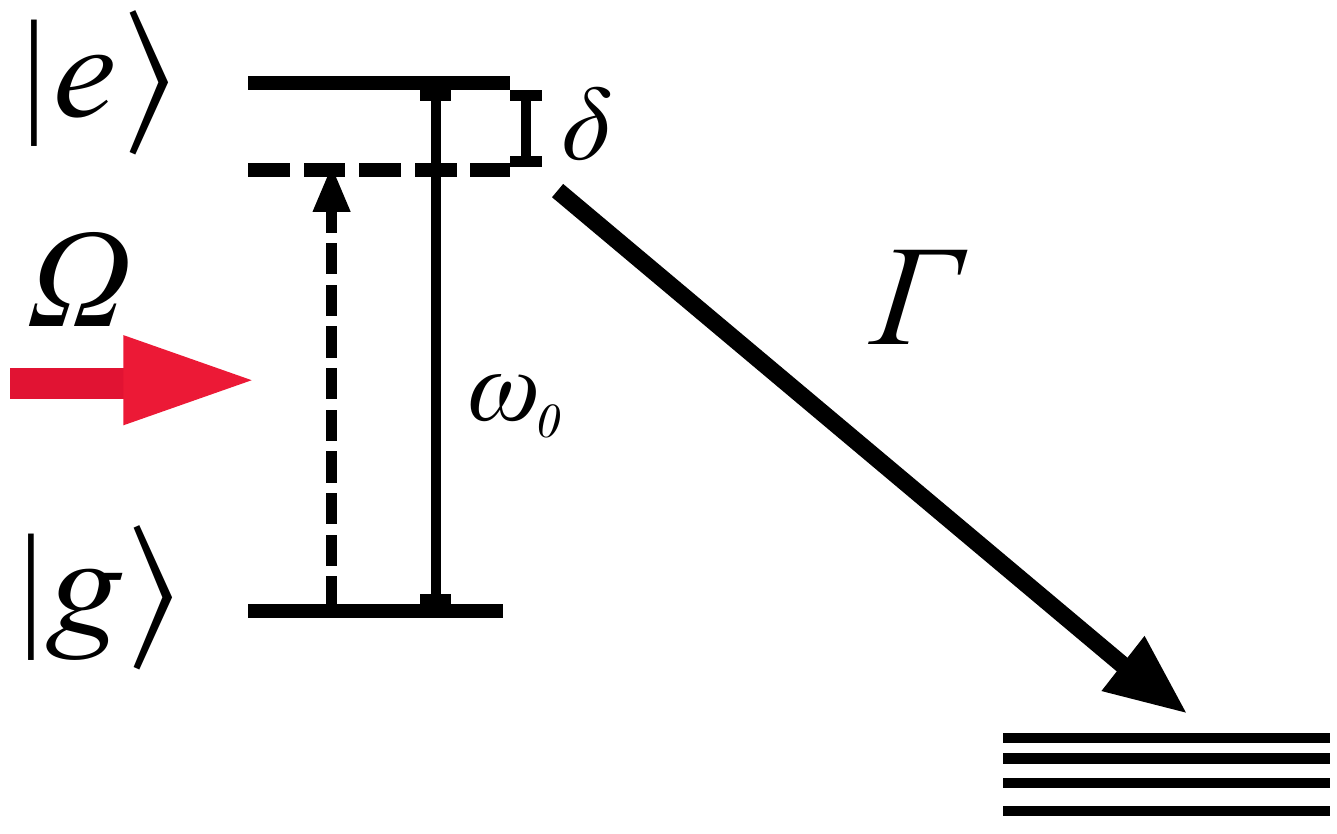}
\caption{\label{figs1}
{\bf Energy-level diagram of an atom.} The excited state $|e\rangle$  has the frequency $\omega_{0}$ relative to the ground state $|g\rangle$ and fast decay modes with the total decay rate $\Gamma$.  A weak near-resonant light with the Rabi frequency $\Omega$ and detuning $\delta$ creates an effective imaginary potential for the ground-state atom, provided that $\Gamma$ is much larger than the spontaneous decay rate from $|e\rangle$ to $|g\rangle$. 
}
\end{figure}

We consider a situation in which atoms in the system have an energy level diagram shown in Supplementary Fig.~S1. Here the excited state $|e\rangle$ has the frequency $\omega_{0}$ relative to the ground state $|g\rangle$ and  fast decay channels to other states with the total decay rate $\Gamma$ much larger than the spontaneous emission rate from $|e\rangle$ to $|g\rangle$. The system is subject to a weak near-resonant light whose electric filed is given by ${\bf E}({\bf x},t)=2{\bf E}_{0}({\bf x})\cos(\omega_{\rm L}t)$. The dynamics of atoms in the levels $\{|g\rangle,|e\rangle\}$ is then described by the many-body Lindblad equation:
\eqn{\label{Master0}
\frac{{\rm d}\hat{\rho}}{{\rm d}t} & = & -\frac{i}{\hbar}[\hat{H},\hat{\rho}]-\frac{\Gamma}{2}\int\left[\hat{\Psi}_{\rm e}^{\dagger}({\bf x})\hat{\Psi}_{\rm e}({\bf x})\hat{\rho}+\hat{\rho}\hat{\Psi}_{\rm e}^{\dagger}({\bf x})\hat{\Psi}_{\rm e}({\bf x})-2\hat{\Psi}_{\rm e}({\bf x})\hat{\rho}\hat{\Psi}_{\rm e}^{\dagger}({\bf x})\right]d{\bf x},
}
where $\hat{\Psi}_{\rm e}$ denotes the field operator of an excited atom and the terms involving $\Gamma$  describe a loss of atoms in the state $|e\rangle$. Here $\hat{H}$ is the Hamiltonian of the interacting two-level atoms:
\begin{equation}
\hat{H}=\hat{\mathcal{H}}_{\rm g}+\hat{\mathcal{H}}_{\rm e}+\hat{{\cal V}}.
\end{equation}
Going onto the rotating frame and making the rotating-wave approximation, the Hamiltonians $\hat{\cal H}_{\rm g}$ and $\hat{\cal H}_{\rm e}$  of ground- and excited-state atoms and the interaction Hamiltonian $\hat{\cal V}$ describing the Rabi coupling between the two atomic levels are given by 
\eqn{\label{Hg}
\hat{{\cal H}}_{\rm g}&=&\int d{\bf x}\left[\hat{\Psi}_{\rm g}^{\dagger}({\bf x})\left(-\frac{\hbar^{2}\nabla^{2}}{2m}+U_{\rm g}({\bf x})\right)\hat{\Psi}_{\rm g}({\bf x})+\frac{g}{2}\hat{\Psi}_{\rm g}^{\dagger}({\bf x})\hat{\Psi}_{\rm g}^{\dagger}({\bf x})\hat{\Psi}_{\rm g}({\bf x})\hat{\Psi}_{\rm g}({\bf x})\right],\\
\hat{{\cal H}}_{\rm e}&=&\int d{\bf x}\;\hat{\Psi}_{\rm e}^{\dagger}({\bf x})\left(-\frac{\hbar^{2}\nabla^{2}}{2m}+U_{\rm e}({\bf x})+\hbar\delta\right)\hat{\Psi}_{\rm e}({\bf x}),\\
\hat{{\cal V}}&=&-\frac{\hbar}{2}\int d{\bf x}\left(\Omega({\bf x})\hat{\Psi}_{\rm g}^{\dagger}({\bf x})\hat{\Psi}_{\rm e}({\bf x})+{\rm H.c.}\right)\equiv\hat{{\cal V}}_{-}+\hat{{\cal V}}_{+},\label{rabi}
}
where $U_{\rm g,e}({\bf x})$'s are optical trapping potentials of the ground- and excited-state atoms created by a far-detuned light, $g$ is the strength of the contact interaction between the ground-state atoms, $\delta=\omega_{\rm L}-\omega_{0}$ is the detuning, $\Omega({\bf x})=2{\bf d}\cdot{\bf E}_{0}(\bf x)/\hbar$ is the Rabi frequency with ${\bf d}=\langle e|\hat{\bf d}|g\rangle$ being the dipole moment, and $\hat{\cal V}_{+(-)}$ are the coupling terms that cause excitation (deexcitation) of the atoms. 
Let us introduce the non-Hermitian Hamiltonian $\hat{\cal H}_{{\rm e},{\rm eff}}$ of the excited-state atoms by
\begin{equation}
\hat{{\cal H}}_{{\rm e},{\rm eff}}=\hat{{\cal H}}_{\rm e}-\frac{i\hbar\Gamma}{2}\int d{\bf x}\hat{\Psi}_{\rm e}^{\dagger}({\bf x})\hat{\Psi}_{\rm e}({\bf x}).
\end{equation}
Then, the time-evolution equation (\ref{Master0}) is written as follows:
\begin{equation}\label{Masterp}
\frac{{\rm d}\hat{\rho}}{{\rm d}t}=-\frac{i}{\hbar}\left[\left(\hat{\mathcal{H}}_{\rm g}+\hat{\mathcal{H}}_{{\rm e},{\rm eff}}+\hat{{\cal V}}\right)\hat{\rho}-\hat{\rho}\left(\hat{\mathcal{H}}_{\rm g}+\hat{\mathcal{H}}_{{\rm e},{\rm eff}}^{\dagger}+\hat{{\cal V}}\right)\right]+\Gamma\int d{\bf x}\hat{\Psi}_{\rm e}({\bf x})\hat{\rho}\hat{\Psi}_{\rm e}^{\dagger}({\bf x}).
\end{equation}
In the limit of rapid decay $\Gamma\gg \delta,\Omega$, we can adiabatically eliminate the rapidly evolving excited states and obtain the effective dynamics of the ground-state atoms. We achieve this by solving  Supplementary Equation (\ref{Masterp}) using the second-order perturbation theory with respect to weak coupling $\hat{\cal V}$  \cite{TK80}. As shown below, the resulting  time-evolution equation for the ground-state atoms is given by Supplementary Equation (\ref{Masterg}), and it reduces to the effective non-Hermitian dynamics   (\ref{postsel}) with the effective Hamiltonian (\ref{nonher}) when the postselection is implemented. 

To perform the perturbative analysis, we work in the interaction picture, where the density matrix is given by
\begin{equation}\label{intpic}
\hat{\tilde{\rho}}_{\rm I}(t)=e^{i\left(\hat{{\cal H}}_{\rm g}+\hat{{\cal H}}_{{\rm e},{\rm eff}}\right)t/\hbar}\hat{\rho}(t)e^{-i\left(\hat{{\cal H}}_{\rm g}+\hat{{\cal H}}_{{\rm e},{\rm eff}}^{\dagger}\right)t/\hbar},
\end{equation}
and a general operator $\hat{\cal O}$ is represented by
\begin{equation}
\hat{{\cal O}}_{{\rm I}}(t)=e^{i\left(\hat{{\cal H}}_{\rm g}+\hat{{\cal H}}_{{\rm e},{\rm eff}}\right)t/\hbar}\hat{{\cal O}}e^{-i\left(\hat{{\cal H}}_{\rm g}+\hat{{\cal H}}_{{\rm e},{\rm eff}}\right)t/\hbar}.
\end{equation}
We note that $\hat{\tilde{\rho}}_{\rm I}$ in Supplementary Equation (\ref{intpic}) is not normalized to unity in general. 
The time-evolution equation (\ref{Masterp}) is then simplified to
\begin{equation}
\dot{\hat{\tilde{\rho}}}_{\rm I}=-\frac{i}{\hbar}\left[\hat{{\cal V}}_{{\rm I}}\hat{\tilde{\rho}}_{\rm I}-\hat{\tilde{\rho}}_{\rm I}\hat{{\cal V}}_{{\rm I}}^{\dagger}\right]+\Gamma\int d{\bf x}\hat{\Psi}_{{\rm I},{\rm e}}({\bf x})\hat{\tilde{\rho}}_{\rm I}\hat{\Psi}_{{\rm I},{\rm e}}^{\dagger}({\bf x}).
\end{equation}
We assume that all the atoms reside in the ground state at $t=0$. Then, we decompose the evolving state $\hat{\tilde{\rho}}_{\rm I}(t)$ into a perturbation series with respect to the weak coupling $\hat{\cal V}_{\rm I}$:
\begin{equation}\label{expansion}
\hat{\tilde{\rho}}_{\rm I}(t)=\hat{\tilde{\rho}}_{\rm I}^{(0)}(t)+\hat{\tilde{\rho}}_{\rm I}^{(1)}(t)+\hat{\tilde{\rho}}_{\rm I}^{(2)}(t)+\cdots,\;\;\;\;\;\left|\hat{\tilde{\rho}}_{\rm I}^{(n)}(t)\right|\propto\left(\frac{|\Omega|}{\Gamma}\right)^{n}\left|\hat{\tilde{\rho}}_{\rm I}^{(0)}(t)\right|,
\end{equation}
where $|\cdots|$ denotes the trace norm. The recursive equations of the first three terms in the expansion (\ref{expansion}) are given by
\eqn{
\dot{\hat{\tilde{\rho}}}_{\rm I}^{(0)} & =&0,\label{zeroth}\\
\dot{\hat{\tilde{\rho}}}_{\rm I}^{(1)} & =&-\frac{i}{\hbar}\left[\hat{{\cal V}}_{{\rm I}}\hat{\tilde{\rho}}_{\rm I}^{(0)}-\hat{\tilde{\rho}}_{\rm I}^{(0)}\hat{{\cal V}}_{{\rm I}}^{\dagger}\right],\label{first}\\
\dot{\hat{\tilde{\rho}}}_{\rm I}^{(2)} & =&-\frac{i}{\hbar}\left[\hat{{\cal V}}_{{\rm I}}\hat{\tilde{\rho}}_{\rm I}^{(1)}-\hat{\tilde{\rho}}_{\rm I}^{(1)}\hat{{\cal V}}_{{\rm I}}^{\dagger}\right]+\Gamma\int d{\bf x}\hat{\Psi}_{{\rm I},{\rm e}}({\bf x})\hat{\tilde{\rho}}_{\rm I}^{(2)}\hat{\Psi}_{{\rm I},{\rm e}}^{\dagger}({\bf x}).\label{second}
}
From Supplementary Equation (\ref{zeroth}), we can take $\hat{\tilde{\rho}}_{\rm I}^{(0)}$ as a time-independent operator. Supplementary Equation (\ref{first}) can formally be integrated to give
\begin{equation}\label{firstsol}
\hat{\tilde{\rho}}_{\rm I}^{(1)}(t)=-\frac{i}{\hbar}\int_{0}^{t}dt'\left[\hat{{\cal V}}_{{\rm I}}(t')\hat{\tilde{\rho}}_{\rm I}^{(0)}-\hat{\tilde{\rho}}_{\rm I}^{(0)}\hat{{\cal V}}_{{\rm I}}^{\dagger}(t')\right].
\end{equation}
To  integrate out the excited states and obtain the effective dynamics of the ground-state atoms, we decompose $\hat{\tilde{\rho}}_{\rm I}^{(2)}$ into  the subspaces of the ground- and excited-state atoms. To do so, we introduce the projection $\hat{\cal P}_{\rm g}$ onto the ground-state manifold by $\hat{{\cal P}}_{\rm g}=\sum_{N}\hat{{\cal P}}_{\rm g}^{N}$, where $\hat{{\cal P}}_{\rm g}^{N}$ denotes the projection onto the subspace spanned by quantum states containing $N$ ground-state atoms only. We also introduce the projection $\hat{\cal Q}_{\rm e}^{1}$ onto quantum states having a single excited-state atom (and an arbitrary number of ground-state atoms). Then, Supplementary Equation (\ref{second}) can be decomposed as
\eqn{\label{secground}
\hat{{\cal P}}_{\rm g}\dot{\hat{\tilde{\rho}}}_{\rm I}^{(2)}\hat{{\cal P}}_{\rm g}&=&-\frac{i}{\hbar}\hat{{\cal P}}_{\rm g}\left[\hat{{\cal V}}_{{\rm I}}\hat{\tilde{\rho}}_{\rm I}^{(1)}-\hat{\tilde{\rho}}_{\rm I}^{(1)}\hat{{\cal V}}_{{\rm I}}^{\dagger}\right]\hat{{\cal P}}_{\rm g}+\Gamma\hat{\cal P}_{\rm g}\int d{\bf x}\hat{\Psi}_{{\rm I},{\rm e}}\hat{{\cal Q}}_{\rm e}^{1}\hat{\tilde{\rho}}_{\rm I}^{(2)}\hat{{\cal Q}}_{\rm e}^{1}\hat{\Psi}_{{\rm I},{\rm e}}^{\dagger}\hat{\cal P}_{\rm g},\\
\hat{{\cal Q}}_{\rm e}^{1}\dot{\hat{\tilde{\rho}}}_{\rm I}^{(2)}\hat{{\cal Q}}_{\rm e}^{1}&=&-\frac{i}{\hbar}\hat{{\cal Q}}_{\rm e}^{1}\left[\hat{{\cal V}}_{{\rm I}}\hat{\tilde{\rho}}_{\rm I}^{(1)}-\hat{\tilde{\rho}}_{\rm I}^{(1)}\hat{{\cal V}}_{{\rm I}}^{\dagger}\right]\hat{{\cal Q}}_{\rm e}^{1},\label{adia}
}
where Supplementary Equation (\ref{adia}) follows from the fact that $\hat{\tilde{\rho}}_{\rm I}^{(2)}$ contains, at most, one excited-state atom. 
We adiabatically eliminate the excited states by integrating out Supplementary Equation (\ref{adia}):
\begin{equation}\label{secondsol}
\hat{{\cal Q}}_{\rm e}^{1}\hat{\tilde{\rho}}_{\rm I}^{(2)}(t)\hat{{\cal Q}}_{\rm e}^{1}=-\frac{i}{\hbar}\hat{{\cal Q}}_{\rm e}^{1}\int_{0}^{t}dt'\left[\hat{{\cal V}}_{{\rm I}}(t')\hat{\tilde{\rho}}_{\rm I}^{(1)}(t')-\hat{\tilde{\rho}}_{\rm I}^{(1)}(t')\hat{{\cal V}}_{{\rm I}}^{\dagger}(t')\right]\hat{{\cal Q}}_{\rm e}^{1}.
\end{equation}
Substituting Supplementary Equations (\ref{firstsol}) and (\ref{secondsol}) into (\ref{secground}), we obtain
\begin{eqnarray}\label{secint}
\hat{{\cal P}}_{\rm g}\dot{\hat{\tilde{\rho}}}_{\rm I}^{(2)}\hat{{\cal P}}_{\rm g} & = & -\frac{1}{\hbar^2}\hat{{\cal P}}_{\rm g}\left[\hat{{\cal V}}_{{\rm I}}(t)\int_{0}^{t}dt'\hat{{\cal V}}_{{\rm I}}(t')\hat{\tilde{\rho}}_{\rm I}^{(0)}+{\rm H.c.}\right]\hat{{\cal P}}_{\rm g}\nonumber\\
 &  & +\frac{\Gamma}{\hbar^{2}}\hat{{\cal P}}_{\rm g}\int d{\bf x}\hat{\Psi}_{{\rm I},{\rm e}}\hat{{\cal Q}}_{\rm e}^{1}\int_{0}^{t}dt'\int_{0}^{t'}dt''\left[\hat{{\cal V}}_{{\rm I}}(t')\hat{\tilde{\rho}}_{\rm I}^{(0)}\hat{{\cal V}}_{{\rm I}}^{\dagger}(t'')+{\rm H.c.}\right]\hat{{\cal Q}}_{\rm e}^{1}\hat{\Psi}_{{\rm I},{\rm e}}^{\dagger}\hat{{\cal P}}_{\rm g}.
\end{eqnarray}
Here, in the second line in Supplementary Equation (\ref{secint}), the terms proportional to $\hat{\cal V}_{\rm I}\hat{\cal V}_{\rm I}\hat{\tilde{\rho}}_{\rm I}^{(0)}$ or $\hat{\tilde{\rho}}_{\rm I}^{(0)}\hat{\cal V}_{\rm I}^{\dagger}\hat{\cal V}_{\rm I}^{\dagger}$ vanish because of the projection $\hat{\cal Q}_{\rm e}^{1}$.
Then, since we assume that the time scale of the strong dissipation is fast compared with other time scales appearing in the system, we approximate  the leading contributions by
\begin{equation}
e^{-i\left(\hat{{\cal H}}_{\rm g}+\hat{{\cal H}}_{{\rm e},{\rm eff}}\right)t/\hbar}\hat{{\cal P}}_{\rm g}\simeq\hat{{\cal P}}_{\rm g},\; \;\;\;e^{-i\left(\hat{{\cal H}}_{\rm g}+\hat{{\cal H}}_{{\rm e},{\rm eff}}\right)t/\hbar}\hat{{\cal Q}}_{\rm e}^{1}\simeq e^{-\Gamma t/2}\hat{{\cal Q}}_{\rm e}^{1}.
\end{equation}
From these equations, it follows that
\eqn{
\hat{\cal P}_{\rm g}\hat{\cal V}_{\rm I}(t)&=&\hat{\cal P}_{\rm g}e^{i\left(\hat{{\cal H}}_{\rm g}+\hat{{\cal H}}_{{\rm e},{\rm eff}}\right)t/\hbar}(\hat{\cal V}_{+}+\hat{\cal V}_{-})e^{-i\left(\hat{{\cal H}}_{\rm g}+\hat{{\cal H}}_{{\rm e},{\rm eff}}\right)t/\hbar}\nonumber\\
&\simeq&\hat{\cal P}_{\rm g}\hat{\cal V}_{-}\hat{\cal Q}_{\rm e}^{1}e^{-i\left(\hat{{\cal H}}_{\rm g}+\hat{{\cal H}}_{{\rm e},{\rm eff}}\right)t/\hbar}\nonumber\\
&\simeq&e^{-\Gamma t/2}\hat{\cal P}_{\rm g}\hat{\cal V}_{-}\hat{\cal Q}_{\rm e}^{1}.\label{ppre1}
}
Similarly, we obtain
\eqn{
\hat{\cal Q}_{\rm e}^{1}\hat{\cal V}_{\rm I}(t)\hat{\cal P}_{\rm g}&=&\hat{\cal Q}_{\rm e}^{1}e^{i\left(\hat{{\cal H}}_{\rm g}+\hat{{\cal H}}_{{\rm e},{\rm eff}}\right)t/\hbar}(\hat{\cal V}_{+}+\hat{\cal V}_{-})e^{-i\left(\hat{{\cal H}}_{\rm g}+\hat{{\cal H}}_{{\rm e},{\rm eff}}\right)t/\hbar}\hat{\cal P}_{\rm g}\nonumber\\
&\simeq& e^{\Gamma t/2}\hat{\cal Q}_{\rm e}^{1}\hat{\cal V}_{+}\hat{\cal P}_{\rm g}\label{ppre2}
}
We then perform the integration in the first line on the right-hand side of Supplementary Equation (\ref{secint}) and obtain
\begin{eqnarray}\label{virtual}
-\frac{1}{\hbar^2}\hat{{\cal P}}_{\rm g}\left[\hat{{\cal V}}_{{\rm I}}(t)\int_{0}^{t}dt'\hat{{\cal V}}_{{\rm I}}(t')\hat{\tilde{\rho}}_{\rm I}^{(0)}+{\rm H.c.}\right]\hat{{\cal P}}_{\rm g} 
&\simeq&-\frac{1}{\hbar^2}\left[\hat{\cal P}_{\rm g}\hat{\cal V}_{-}\hat{\cal Q}_{\rm e}^{1}e^{-\Gamma t/2}\int_{0}^{t}dt' e^{\Gamma t'/2}\hat{\cal Q}_{\rm e}^{1}\hat{\cal V}_{+}\hat{\cal P}_{\rm g}\hat{\tilde{\rho}}_{\rm I}^{(0)}+{\rm H.c.}\right]\nonumber\\ 
& \simeq & -\frac{2}{\hbar^2\Gamma}\left(\hat{{\cal P}}_{\rm g}\hat{{\cal V}}_{-}\hat{{\cal V}}_{+}\hat{{\cal P}}_{\rm g}\hat{\tilde{\rho}}_{\rm I}^{(0)}+\hat{\tilde{\rho}}_{\rm I}^{(0)}\hat{{\cal P}}_{\rm g}\hat{{\cal V}}_{-}\hat{{\cal V}}_{+}\hat{{\cal P}}_{\rm g}\right)\nonumber\\
 & = & -\left\{ \int d{\bf x}\frac{\left|\Omega({\bf x})\right|^{2}}{2\Gamma}\hat{\Psi}_{\rm g}^{\dagger}({\bf x})\hat{\Psi}_{\rm g}({\bf x}),\hat{\tilde{\rho}}_{\rm I}^{(0)}\right\}, \label{secpre1}
\end{eqnarray}
where we use Supplementary Equations (\ref{ppre1}), (\ref{ppre2}), and the relations $\hat{\cal P}_{\rm g}\hat{\tilde{\rho}}_{\rm I}^{(0)}\hat{\cal P}_{\rm g}=\hat{\tilde{\rho}}_{\rm I}^{(0)}$ and $(\hat{\cal Q}_{\rm e}^{1})^2=\hat{\cal Q}_{\rm e}^{1}$ in the first line, and use Supplementary Equation (\ref{rabi}) to derive the last line. To calculate the last line in Supplementary Equation (\ref{secint}), we approximate
\begin{eqnarray}
\hat{{\cal Q}}_{\rm e}^{1}\int_{0}^{t}dt'\int_{0}^{t'}dt''\left[\hat{{\cal V}}_{{\rm I}}(t')\hat{\tilde{\rho}}_{\rm I}^{(0)}\hat{{\cal V}}_{{\rm I}}^{\dagger}(t'')+{\rm H.c.}\right]\hat{{\cal Q}}_{\rm e}^{1} & \simeq& 2\int_{0}^{t}dt'\int_{0}^{t'}dt''e^{\Gamma(t'+t'')/2} \hat{{\cal Q}}_{\rm e}^{1}\hat{{\cal V}}_{+}\hat{\tilde{\rho}}_{\rm I}^{(0)}\hat{{\cal V}}_{-}\hat{{\cal Q}}_{\rm e}^{1}\nonumber\\ 
 & \simeq & \frac{4e^{\Gamma t}}{\Gamma^{2}}\hat{{\cal V}}_{+}\hat{\tilde{\rho}}_{\rm I}^{(0)}\hat{{\cal V}}_{-}, \label{intpre1}
\end{eqnarray}
and
\begin{equation}\label{intpreee2}
\hat{{\cal P}}_{\rm g}\hat{\Psi}_{{\rm I},{\rm e}}\hat{{\cal Q}}_{\rm e}^{1}\simeq e^{-\Gamma t/2}\hat{{\cal P}}_{\rm g}\hat{\Psi}_{\rm e}\hat{{\cal Q}}_{\rm e}^{1},\;\;\hat{{\cal Q}}_{\rm e}^{1}\hat{\Psi}_{{\rm I},{\rm e}}^{\dagger}\hat{{\cal P}}_{\rm g}\simeq e^{-\Gamma t/2}\hat{{\cal Q}}_{\rm e}^{1}\hat{\Psi}_{\rm e}^{\dagger}\hat{{\cal P}}_{\rm g}
\end{equation}
The last line in Supplementary Equation (\ref{secint}) can then be calculated as 
\eqn{
&&\frac{\Gamma}{\hbar^{2}}\hat{{\cal P}}_{\rm g}\int d{\bf x}\hat{\Psi}_{{\rm I},{\rm e}}\hat{{\cal Q}}_{\rm e}^{1}\int_{0}^{t}dt'\int_{0}^{t'}dt''\left[\hat{{\cal V}}_{{\rm I}}(t')\hat{\tilde{\rho}}_{\rm I}^{(0)}\hat{{\cal V}}_{{\rm I}}^{\dagger}(t'')+{\rm H.c.}\right]\hat{{\cal Q}}_{\rm e}^{1}\hat{\Psi}_{{\rm I},{\rm e}}^{\dagger}\hat{{\cal P}}_{\rm g}\nonumber\\
&\simeq&\frac{4}{\hbar^2\Gamma}\int d{\bf x}\hat{\cal P}_{\rm g}\hat{\Psi}_{\rm e}({\bf x})\hat{\cal V}_{+}\hat{\cal P}_{\rm g}\hat{\tilde{\rho}}_{\rm I}^{(0)}\hat{\cal P}_{\rm g}\hat{\cal V}_{-}\hat{\Psi}_{\rm e}^{\dagger}({\bf x})\hat{\cal P}_{\rm g}\nonumber
\\
&\simeq&\hat{\cal P}_{\rm g}\int d{\bf x}\frac{\left|\Omega({\bf x})\right|^{2}}{\Gamma}\hat{\Psi}_{\rm g}({\bf x})\hat{\tilde{\rho}}_{\rm I}^{(0)}\hat{\Psi}_{\rm g}^{\dagger}({\bf x})\hat{\cal P}_{\rm g},\label{secpre2}
}
where we use Supplementary Equations  (\ref{intpre1}) and (\ref{intpreee2}) and $\hat{\cal P}_{\rm g}\hat{\tilde{\rho}}_{\rm I}^{(0)}\hat{\cal P}_{\rm g}=\hat{\tilde{\rho}}_{\rm I}^{(0)}$ in the second line. To derive the last line, we use the following relation
\eqn{\hat{{\cal P}}_{\rm g}\hat{\Psi}_{\rm e}({\bf x})\hat{{\cal V}}_{+}\hat{{\cal P}}_{\rm g}=-\frac{\hbar\Omega^{*}({\bf x})}{2}\hat{{\cal P}}_{\rm g}\hat{\Psi}_{\rm g}({\bf x})\hat{{\cal P}}_{\rm g}.
}

From equations (\ref{secint}), (\ref{secpre1}), (\ref{secpre2}) and $\hat{\cal P}_{\rm g}\dot{\hat{\tilde{\rho}}}_{\rm I}^{(0)}\hat{\cal P}_{\rm g}=\hat{\cal P}_{\rm g}\dot{\hat{\tilde{\rho}}}_{\rm I}^{(1)}\hat{\cal P}_{\rm g}=0$, the effective time-evolution equation of the ground-state atoms is obtained as
\eqn{\label{Masterg}
\frac{{\rm d}\hat{\rho}_{\rm g}}{{\rm d}t}&=&-\frac{i}{\hbar}\left(\hat{{\cal H}}_{{\rm g},{\rm eff}}\hat{\rho}_{\rm g}-\hat{\rho}_{\rm g}\hat{{\cal H}}_{{\rm g},{\rm eff}}^{\dagger}\right)+\int d{\bf x}\frac{\left|\Omega({\bf x})\right|^{2}}{\Gamma}\hat{\Psi}_{\rm g}({\bf x})\hat{\rho}_{\rm g}\hat{\Psi}_{\rm g}^{\dagger}({\bf x}),\\
\hat{{\cal H}}_{{\rm g},{\rm eff}}&\equiv&\hat{{\cal H}}_{\rm g}-i\hbar\int d{\bf x}\frac{|\Omega({\bf x})|^{2}}{2\Gamma}\hat{\Psi}_{\rm g}^{\dagger}({\bf x})\hat{\Psi}_{\rm g}({\bf x}),\label{nonher}
}
where we go back to the Schr{\"o}dinger picture and introduce the density matrix $\hat{\rho}_{\rm g}$ projected onto the ground-state manifold by
\begin{equation}
\hat{\rho}_{\rm g}(t)=\hat{{\cal P}}_{\rm g}\hat{\rho}(t)\hat{{\cal P}}_{\rm g}\simeq\hat{{\cal P}}_{\rm g}\left(\hat{\rho}^{(0)}(t)+\hat{\rho}^{(1)}(t)+\hat{\rho}^{(2)}(t)\right)\hat{{\cal P}}_{\rm g}.
\end{equation}
The non-Hermitian Hamiltonian  (\ref{nonher}) describes the effective dynamics of the system when we postselect realizations in which no quantum jumps occur, i.e., no atoms escape from the ground state  \cite{HC93,AD15,YA16}. To clarify this point, let us assume that $N$ ground-state atoms are initially prepared, i.e., $\hat{\cal P}_{\rm g}^{N}\hat{\rho}(0)\hat{\cal P}_{\rm g}^{N}=\hat{\rho}(0)$.  This initial condition implies
\begin{equation}
\hat{{\cal P}}_{\rm g}^{N+l}\hat{\rho}_{\rm g}(0)\hat{{\cal P}}_{\rm g}^{N+l}=0\;\;\;\;\;(l=1,2,\ldots),\label{initialc}
\end{equation}
where we use $\left[\hat{{\cal P}}_{\rm g}^{N},\hat{{\cal P}}_{\rm g}\right]=0$.
From  Supplementary Equations (\ref{Masterg}) and (\ref{initialc}), we can in particular show that, during the course of the time evolution, 
\eqn{\hat{{\cal P}}_{\rm g}^{N+1}\hat{\rho}_{\rm g}(t)\hat{{\cal P}}_{\rm g}^{N+1}=0.\label{initialc2}}
Let us now consider the dynamics of the postselected system $\hat{\tilde{\rho}}_{\rm g}^{N}(t)\equiv\hat{{\cal P}}_{\rm g}^{N}\hat{\rho}(t)\hat{{\cal P}}_{\rm g}^{N}=\hat{{\cal P}}_{\rm g}^{N}\hat{\rho}_{\rm g}(t)\hat{{\cal P}}_{\rm g}^{N}$, where the dynamics is conditioned such that no atoms are lost from the initial state. Using Supplementary Equations (\ref{Masterg}) and (\ref{initialc2}), we can show that the dynamics of the postselected system $\hat{\tilde{\rho}}^{N}_{\rm g}$ is governed by the non-Hermitian Hamiltonian (\ref{nonher}):
\begin{equation}\label{postsel}
\frac{{\rm d}\hat{\tilde{\rho}}_{\rm g}^{N}}{{\rm d}t}=-\frac{i}{\hbar}\left(\hat{{\cal H}}_{{\rm g},{\rm eff}}\hat{\tilde{\rho}}_{\rm g}^{N}-\hat{\tilde{\rho}}_{\rm g}^{N}\hat{{\cal H}}_{{\rm g},{\rm eff}}^{\dagger}\right).
\end{equation}
Some remarks are in order here.
First, an imaginary potential  $-i|\Omega({\bf x})|^2/(2\Gamma)$ in Supplementary Equation (\ref{nonher}) arises from the second-order process of a virtual excitation and de-excitation of the ground-state atoms (see Supplementary Equation (\ref{virtual})). Since no atoms are lost in this process, the non-Hermitian contribution exists even when we do not observe actual losses of atoms.  Physically, such a contribution originates from the measurement backaction associated with continuous monitoring of the population of atoms in the excited state  \cite{HC93}. 
Second, we note that the expression of the imaginary potential indicates that the loss rate of atoms from the ground state is suppressed by a factor of $\Omega/\Gamma$ for large $\Gamma$. In particular, in the limit of $\Gamma\to\infty$, the dynamics reduces to the Hermitian evolution governed by $\hat{\cal H}_{\rm g}$. This limit can be  interpreted as the quantum Zeno dynamics, where the strong measurement confines the dynamics into the decay-free subspace and the time-evolution obeys the effective ``Zeno" Hamiltonian. In our model, such a Hamiltonian is given by $\hat{\cal H}_{\rm g}=\hat{\cal P}_{\rm g}\hat{H}\hat{\cal P}_{\rm g}$, where the total Hamiltonian $\hat{H}$ is projected onto the decay-free, ground-state manifold. In a general case of a strong but finite $\Gamma$, we need to perform careful perturbative analyses to obtain the correction terms beyond the quantum Zeno dynamics, as we have conducted above. 

We are now in a position to derive the PT-symmetric Hamiltonian. We consider a system confined in a one-dimensional (1D) optical trap, and assume that the system is subject to a real shallow periodic potential (optical lattice), $U(x)=V_{\rm r}\cos(2\pi x/d)$, which can be created by a far-detuned off-resonant light. Here $V_{\rm r}$ is the lattice potential which is controlled by changing the intensity of the light, and $d=\lambda/2$ is the lattice spacing. We then superimpose the near-resonant standing-wave light discussed above with the displacement of $d/4$. We thus have ${\bf E}_{0}(x)=\mathcal{E}_{0}\cos(kx-\pi/4)$ with $k=2\pi/\lambda$. From Supplementary Equations (\ref{Hg}) and (\ref{nonher}), the resulting Hamiltonian is obtained as
\eqn{
\hat{\cal H}_{{\rm eff}} = \int dx\hat{\Psi}^{\dagger}(x)\left(-\frac{\hbar^{2}\nabla^{2}}{2m}+V(x)\right)\hat{\Psi}(x)+\frac{g}{2}\int dx\hat{\Psi}^{\dagger}(x)\hat{\Psi}^{\dagger}(x)\hat{\Psi}(x)\hat{\Psi}(x),
\label{pre1}
}
where we drop the subscript $\rm g$ and introduce
\eqn{
V(x) & = & V_{\rm r}\cos\left(\frac{2\pi x}{d}\right)-iV_{\rm i}\sin\left(\frac{2\pi x}{d}\right),\label{potential}\\
V_{\rm i} & = & \frac{|d|^{2}\mathcal{E}_{0}^{2}}{\hbar\Gamma }.
}
Here we redefine the interaction parameter $g$ by incorporating the renormalization factor due to the 1D confinement and ignore the constant term $-iV_{\rm i}N$ proportional to the total number of atoms. This constant term is irrelevant in the postselected dynamics here because it is cancelled upon the normalization of the quantum state. The effective Hamiltonian (\ref{pre1}) satisfies PT symmetry because the potential satisfies the condition $V(x)=V^{*}(-x)$. While we here assume that the wavelengths of the lasers creating the real and imaginary potentials in Supplementary Equation (\ref{potential}) are the same, this assumption can be well  met because the detuning required for the real potential causes only a negligible shift in $\lambda$, as detailed later. 

%because it cancels out when we take the normalization of the quantum state, $|\Psi_{t}\rangle=\exp(-i\hat{H}_{\rm eff}t/\hbar)|\Psi_{0}\rangle/\|\exp(-i\hat{H}_{\rm eff}t/\hbar)|\Psi_{0}\rangle\|$

Finally, we derive the low-energy effective theory. 
An interacting 1D Bose gas as described by Supplementary Equation (\ref{pre1}) without the potential $V(x)$ is described  at low energies by the Tomonaga-Luttinger liquid  theory  \cite{TG03}:
\eqn{\label{H}
\hat{H}_{\rm TLL}=\int dx\frac{\hbar v}{2\pi}\left[K(\partial_{x}\hat{\theta})^2+\frac{1}{K}(\partial_{x}\hat{\phi})^2\right].
}
Here $\hat{\theta}$ is related to the phase of the bosonic field operator $\hat{\Psi}^{\dagger}(x)=\sqrt{\hat{\rho}(x)}e^{-i\hat{\theta}(x)}$, and $\hat{\phi}$ is related to the density operator as
\eqn{\label{rho}
\hat{\rho}(x)=\left[\rho_{0}-\frac{1}{\pi}\partial_{x}\hat{\phi}\right]\sum_{p=-\infty}^{\infty}e^{2ip(\pi\rho_{0}x-\hat{\phi}(x))},
}
where $\rho_{0}$ is the average atomic density.
We discuss the perturbative role of the potential term
\eqn{\label{pre2}
\hat{H}_{V}=\int dx\;V(x)\hat{\Psi}^{\dagger}(x)\hat{\Psi}(x)=\int dx\;V(x)\hat{\rho}(x).
} 
Since we are interested in the commensurate phase transition, we assume the unit filling $\rho_{0}d=1$, i.e., one atom per site.  By substituting Supplementary Equations (\ref{potential}) and (\ref{rho}) into Supplementary Equation (\ref{pre2}) and ignoring fast oscillating terms, we obtain 
\eqn{
\hat{H}_{V}=\rho_{0}V_{\rm r}\int dx\cos\left[2\hat{\phi}(x)\right]-i\rho_{0}V_{\rm i}\int dx\sin\left[2\hat{\phi}(x)\right].\label{efft}
}
Defining $\alpha_{\rm r,i}\equiv\pi\rho_{0}V_{\rm r,i}$, we arrive at the PT-symmetric potential term in equation (2) in the main text.
\\
\\
{\bf Supplementary Note 2 - Experimental implementation and signatures in ultracold atoms}
\\
We here describe detailed experimental signatures of the PT-symmetric system in ultracold atoms. To create an imaginary optical potential, we need to realize atomic levels as illustrated in Supplementary Fig.~S1. The fast decay modes can be realized by choosing (i) appropriate spontaneous emission processes or (ii) light-induced transitions. In the scheme (i), one can use the $F=3$ to $F'=3$ transition ($5S_{1/2}$ to $5P_{3/2}$) of $^{85}{\rm Rb}$ atoms to create an imaginary potential   \cite{AT03}, where the excited $F'=3$ state has a decay channel to the $F=2$ state. Implementations of complex potentials have also been demonstrated by using other metastable atomic states  \cite{KSJ98,MKO99,RS05}. The postselection of the null measurement outcomes can be implemented by, e.g., applying the state-selective imaging technique  \cite{FT15}. Here we first load the ground-state atoms into a real optical potential with an accurately estimated  number of atoms. Such a preparation has been achieved with the single-site addressing technique in ultracold atomic experiments. Then we ramp up an imaginary potential and let the system evolve in time. When an atom is excited by a weak near-resonant light, it quickly decays into modes other than the original ground state. Thus, the postselection can be realized by applying the state-selective imaging at the final stage of the time evolution, thereby measuring the number of atoms residing in the ground state, and selecting the realizations in which this number is unchanged between the initial and final states. In this way, we can postselect processes with null quantum jump. We note that the experimental fidelity of measuring the atom number with such site-resolved imaging has reached almost unit fidelity. In view of this development, we expect that the postselection process as described above can be performed with  near-unit fidelity. We note that  various types of postselections have already been achieved owing to the high experimental fidelity  \cite{FT15}.

We next discuss the scheme (ii) that exploits light-induced transitions. Here, the level structures for creating an imaginary potential can be obtained by inducing a fluorescent transition between the excited state and a state other than the original ground state. In this setup, when the ground-state atom is excited, it is quickly lost from an optical potential because the recoil energy due to fluorescent imaging light causes heating of atoms. However,  only a few tens of scattered photons  are enough for the loss of those atoms and thus creating an imaginary potential. If the resulting loss rate $\Gamma$ is much larger than both the spontaneous emission rate and the Rabi frequency $\Omega$, we can adiabatically eliminate the excited state and implement an effective imaginary potential  \cite{MKO99}. The postselection of the null outcome can be realized, for example, by continuously monitoring the system with quantum gas microscopy. If an atom is excited during the time evolution, it emits photons which can be detected by quantum gas microscopy  \cite{BWS09}. Thus, by selecting the events in which no atoms emit photons, the postselection of null quantum jumps (no atomic loss) can be realized. We may also double check the postselection process by performing a projective atom-number measurement at the final stage of the time evolution and ensuring that the atom number stays constant during the time evolution, as described above.

While substantial atomic losses usually lead to experimental difficulties, our theoretical predictions are accessible by using a very weak imaginary potential with which the atomic loss rate can be made  arbitrarily small. This is because the key parameter which drives the phase transitions discussed in the main text is the ratio between $g_{\rm i}$ and $g_{\rm r}$,  which is equivalent to the ratio between the amplitudes of the  imaginary and real potentials (see Supplementary Equation (\ref{efft})), and the imaginary potential required to induce the transition can be made very weak if the depth of the real part of the optical potential is chosen to be sufficiently small. Indeed, such a weak imaginary potential can be implemented in our model since the atomic loss rate is suppressed by a factor of $\Omega/\Gamma$ in the limit of large $\Gamma$  (see Supplementary Equation (\ref{nonher})). 

Since the depth of the real potential can  be made small, and the condition on the Rabi frequency $\Delta_{\rm off}>\Omega$ can easily be met owing to the smallness of the optical depth, the only requirement for the detuning $\Delta_{\rm off}$ of the off-resonant light for the real part of the optical potential is $\Delta_{\rm off}>\Gamma$. This point validates our assumption that the real and imaginary potentials have the same periodicity. For example, for (i) the spontaneous emission process in $^{85}$Rb or (ii) the light-induced transition in the $D2$ transition of $^{87}$Rb, $\Gamma$ is of the order of tens of MHz  \cite{AT03}. Thus, if we set the detuning at $\Delta_{\rm off}=100\;{\rm GHz}$, the off-resonant condition is well satisfied, while such a  detuning causes a less than $0.1$\% shift in the optical wavelength.  
 
We finally discuss experimental signatures of the theoretical predictions described in the main text. First, the measurement-induced Berezinskii-Kosterlitz-Thouless (BKT) transition corresponds to a 1D  superfluid-to-Mott-insulator transition for ultracold atoms. This is associated with a power-law divergence in the momentum distribution at zero momentum, which can be detected by applying the standard techniques such as time-of-flight imaging. Second, the PT symmetry breaking can be probed by detecting the single-mode lasing dynamics of the system. In the PT-broken region, the system has an excited state whose eigenvalue possesses a positive imaginary contribution; such an excited state would have an exponentially growing amplitude in the time evolution. Thus, after exciting the system through, e.g., shaking of an optical lattice, the system eventually approaches  the state having the largest imaginary part of the eigenvalue. Such a single-mode lasing dynamics  entails a significant decrease in the entropy of the system, which can be probed from shot-to-shot fluctuations in $\it in$-$\it situ$ imaging of atomic gases  \cite{EM11,FT15}. Third, the anomalous variation of the critical exponent shown in Fig.~3 in the main text can be investigated through the analysis of the shot-to-shot noise correlations in density fluctuations of a 1D Bose gas. 
\begin{figure}[h!]
\includegraphics[width=150mm]{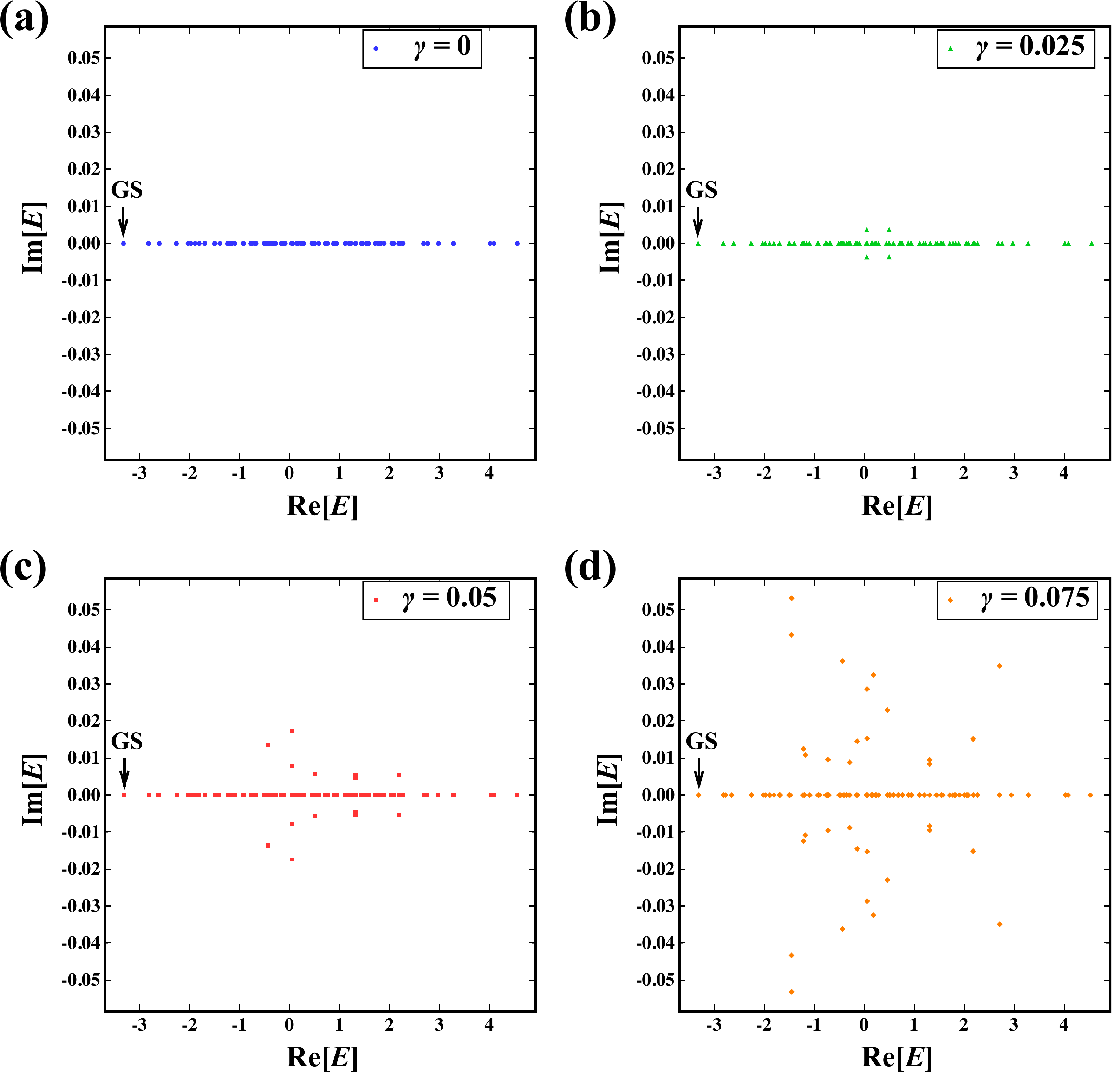}
 \caption{\label{figs3}
 {\bf Exact finite-size spectra.} The spectra of the lattice spin-chain model are plotted with the parameters $-\Delta=0.5,h_{s}=0.0,N=12$ for different strengths of the imaginary hopping (a) $\gamma=0$, (b) $\gamma=0.025$, (c) $\gamma=0.05$, and (d) $\gamma=0.075$. When the PT symmetry is broken, some of excited states have pairs of complex eigenvalues which are conjugate to each other, while the ground state remains to have a real eigenvalue. The plotted energy levels reside in the sector $(S^z =0,q=0,P=T=1)$. The ground state (GS) is indicated by the black arrow. 
}
\end{figure}
\\
\\
{\bf Supplementary Note 3 - Preparation of the ground state in the PT-symmetric spin-chain model}
\\
Let us here discuss how we can study the ground state of the PT-symmetric spin-chain model in the PT-broken regime. When the PT symmetry is broken, some excited eigenstates turn out to have complex pairs of eigenvalues while the ground state remains to have a real eigenvalue (see Supplementary Fig.~S2(a)-(d) for typical spectra). In particular, there exist high-lying unstable modes having positive imaginary parts of eigenvalues. As a result, if the system is significantly perturbed and highly excited, the amplitudes of these modes can grow in time and eventually govern the physical properties of the system.  This is reminiscent of the phenomenon known as parametric instability or self-pulsing  \cite{DS08} in excition-polariton systems, which in general destroys the off-diagonal quasi-long-range order in 1D Bose systems  \cite{IC05}.

In contrast, our main focus here is on the ground state that sustains the quantum critical behavior. This state is indeed relevant in our setup, where the system is initially prepared in the zero-temperature state of the hermitian Hamiltonian and then the imaginary part of the potential required for the PT symmetry is adiabatically ramped up. We numerically demonstrate in Supplementary Fig.~S3 that the system remains in the ground state with almost unit fidelity for a long time interval. Here we consider the spin-chain model (equation (4) in the main text) and adiabatically ramp up the imaginary term with the time dependence $\gamma(t)=\gamma\times\left(1-2/\left(e^{(t/\tau)^2}+1\right)\right)$, where $\tau$ characterizes the timescale of the operation. The initial state $|\Psi(0)\rangle$ is chosen to be the ground state of the Hamiltonian with $\gamma(0)=0$, and the time evolution $|\Psi(t)\rangle$ is calculated by diagonalizing the Hamiltonian at each time step. Supplementary Figure~S3 shows the ground-state fidelity $|\langle\Psi_{{\rm GS},\gamma(t)}|\Psi(t)\rangle|$ of the instantaneous Hamiltonian with $\gamma(t)$, indicating that the system remains in the ground state for a time much longer than the ramping time $\tau$. Using a typical experimental time scale $\hbar/J=3.6/(2\pi)\:{\rm ms}$, the lifetime of the ground state can reach $\sim150\:{\rm ms}$, which is sufficiently long compared with a typical operation time of ultracold atom experiments. We note that the first signature of the enhancement of superfluid correlation can appear from a relatively small size such as $\sim 10$ sites (see Fig. 3b in the main text).
\begin{figure}[t]
\includegraphics[width=120mm]{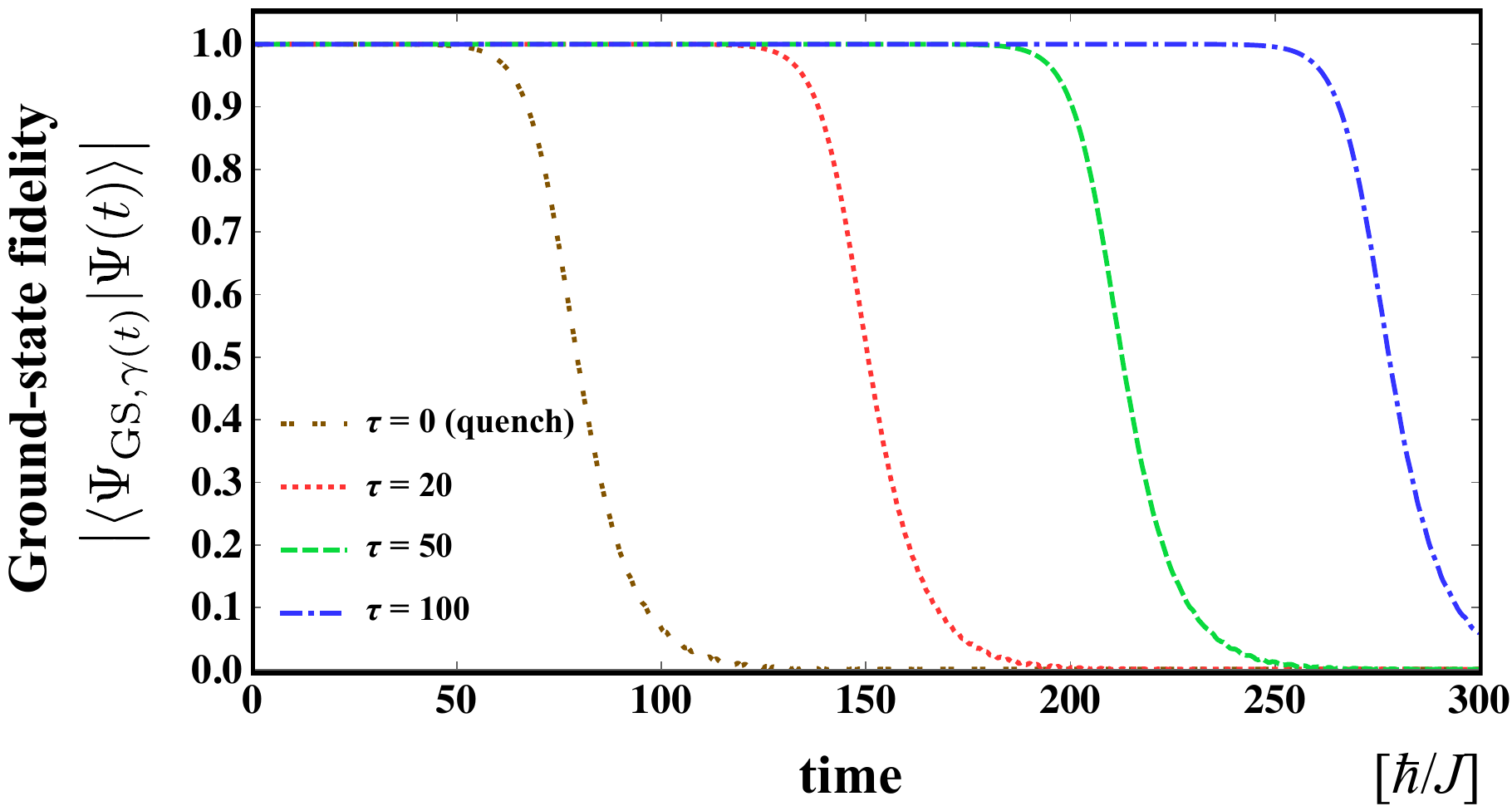}
 \caption{\label{figs4}
{\bf Ground-state fidelity in the PT-broken regime.} The time evolution of the ground-state fidelity of the system in the PT-broken regime is plotted for several different values of $\tau$. The imaginary hopping term $\gamma$ is ramped up with different timescales $\tau=0,20,50,100$. The ground state $|\Psi_{{\rm GS},\gamma(t)}\rangle$ is calculated from the exact diagonalization of the Hamiltonian at each time step. The parameters are set to $-\Delta=0.5$, $h_{s}=0$, $\gamma=0.05$, and $N=12$. 
}
\end{figure}
\clearpage
\noindent
{\bf Supplementary Methods - Determination of the phase diagram of the lattice model}
\\
Here we describe in detail the methods for determining the BKT and PT phase boundaries in the lattice model (equation (4) in the main text). We first describe the so-called level spectroscopy method  \cite{KN95}, which has been developed to accurately determine the BKT transition point. The key idea of this method is to relate the low-energy spectrum to the running coupling constants that appear in the renormalization group (RG) equations. Under the periodic boundary condition, the lattice Hamiltonian  is invariant with respect to spin rotation about the $z$ axis,  translation by two sites, space inversion, and  spin reversal. The corresponding conserved quantum numbers are the total magnetization $S^{z}\equiv\sum_{i=1}^{N}S_{\rm i}^{z}$, the wavenumber $q=2\pi k/L$ ($k\in \mathbb{Z},L\equiv N/2$), the parity $P=\pm 1$, and the spin reversal $T=\pm 1$. The ground state with  energy $E_{\rm g}(L)$ resides in the sector $(S^{z}=0,q=0,P=T=1)$. We denote the second lowest eigenenergy in this sector by $E_0$ and the lowest eigenenergy in the sector $(S^{z}=\pm 4,q=0,P=1)$ by $E_3$. Near the BKT transition line, these excitation energies satisfy  \cite{KN95}
\eqn{
E_{0}(L)-E_{\rm g}(L)&=&\frac{2\pi v}{L}\left(2+\frac{1}{3}\delta(l)-\frac{8}{3}g'(l)\right),\\
E_{3}(L)-E_{\rm g}(L)&=&\frac{2\pi v}{L}(2-\delta(l)),
}
where $\delta\equiv K-2$, $g'\equiv\sqrt{g_{\rm r}^{2}-g_{\rm i}^{2}}$, and the logarithmic RG scale $l$ is related to the system size $L$ via $e^{l}=L/\pi$. At the lowest order of the RG flow equation (3) in the main text, the boundary of the BKT transition corresponds to $\delta=2g'$. Since $E_{0}=E_{3}$ is equivalent to this condition,  the BKT transition point is determined from the crossing point of these two energy levels. In our model, this corresponds to the crossing of the levels shown as the red dashed line and the blue solid line in Supplementary Fig.~S4. In numerical calculations, we obtain the excitation energy of the energy level $(S^{z}=\pm 4,q=0,P=1)$ by multiplying that of the level $(S^{z}=\pm1,q=0,P=1)$ by a factor of 16 to minimize a change in the field-theory parameters due to an increase in the total magnetization $S^z$ in finite-size systems. We note that, even though we consider a non-Hermitian model here, the level spectroscopy method is applicable because the BKT phase boundary is entirely within the PT-unbroken region and the low-energy spectrum is thus equivalent to that of the sine-Gordon model as proved in the main text. We calculate the transition point for different system sizes (Supplementary Fig.~S4(a)-(c)), and  extrapolate it to the thermodynamic limit to determine the BKT transition point (Supplementary Fig.~S4(d)). Since $-\Delta = \cos (\pi / 2K)$ for $h_s=\gamma=0$ and the BKT transition occurs near $K=2$, our analysis focuses on a region around $-\Delta=\cos(\pi/4)=1/\sqrt{2}$. 

The PT threshold is determined from the first coalescence point in the low-energy spectrum. To confirm that the identified point indeed represents an exceptional point of the spectrum,  we plot the square of the energy difference $(\delta E)^2$ and test the square-root scaling of $\delta E$ which appears when an exceptional point is formed by the coalescence of two eigenstates  \cite{TK80} (see insets in Supplementary Fig.~S4(a)-(c)). We then perform a linear fit to the $(\delta E)^2$-$\gamma$ plot and identify the PT threshold $\gamma_{PT}$ for different system sizes. Finally, we  extrapolate it to the thermodynamic limit and determine the PT symmetry breaking point (Supplementary Fig.~S4(d)).

\clearpage

\begin{figure}[h!]
\includegraphics[width=150mm]{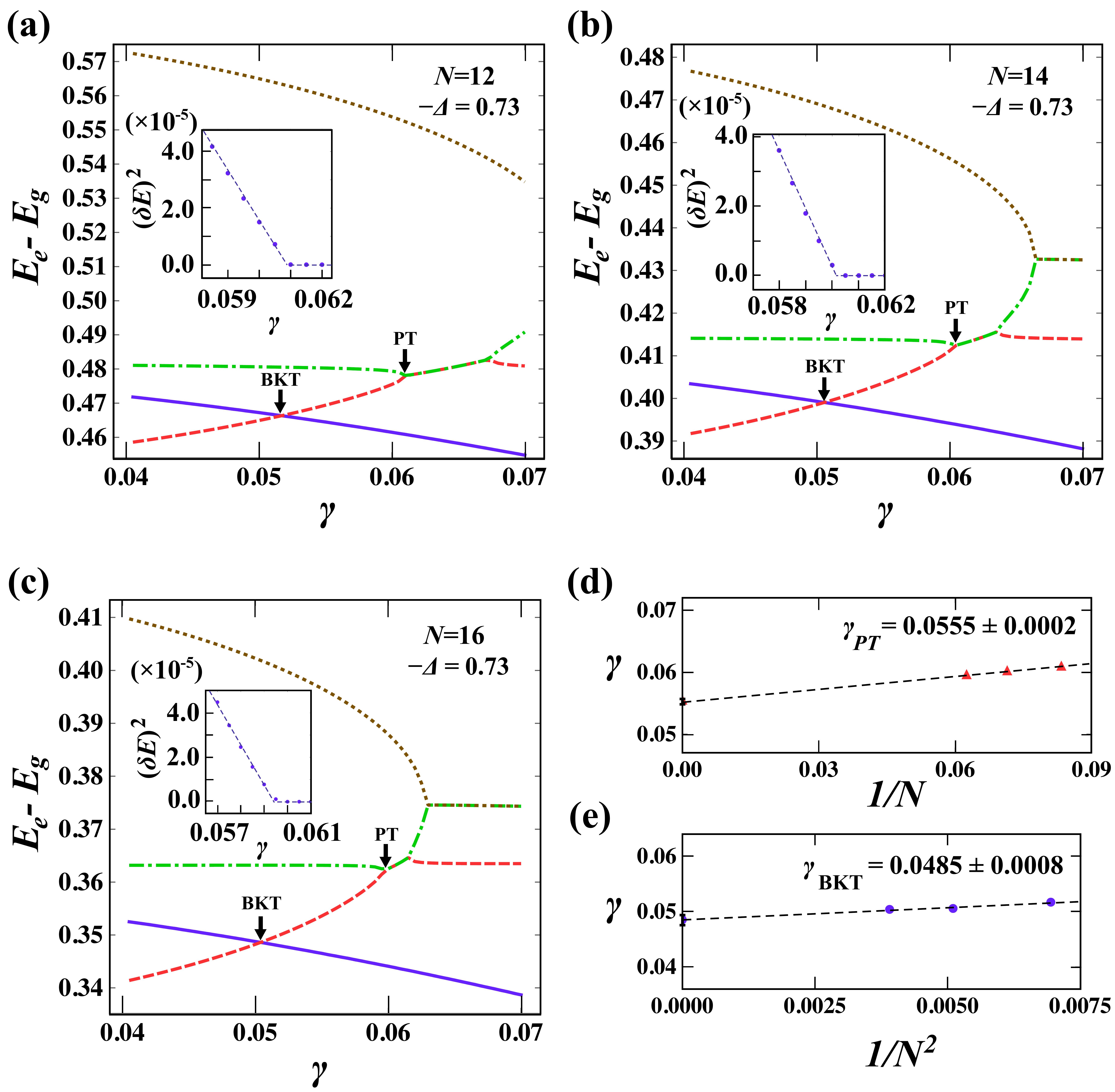}
 \caption{\label{figs2}
 {\bf Exact finite-size spectra for different system sizes.} The spectra are plotted with the parameters $-\Delta=0.73$ and $h_{s}=0.1$ for different system sizes (a) $N=12$, (b) $N=14$, and (c) $N=16$. Here the three lowest excited levels in the $(S^{z}=0,q=0,P=T=1)$ sector (red, green, and yellow curves from the lowest), and the lowest excitation energy in the $(S^{z}=\pm4, q=0, P=1)$ sector (blue curve) are plotted.
 The Berezinskii-Kosterlitz-Thouless (BKT) transition point corresponds to the crossing point of the two energy levels in $(S^{z}=0,q=0,P=T=1)$ (red) and $(S^{z}=\pm4, q=0, P=1)$ (blue). The PT transition point corresponds to the first coalescence point of two low-energy levels (e.g., red and green), which is confirmed to be an exceptional point of the spectrum by testing the square-root scaling of the energy difference $\delta E$ between the two coalescing levels (inset).  (d) The PT threshold ($\gamma_{PT}$) and (e) the BKT transition point ($\gamma_{\rm BKT}$) are determined by extrapolating finite-size data to the thermodynamic limit. 
}
\end{figure}

\end{document}